**TITLE**

Adaptive cognition implemented with a context-aware and flexible neuron for next-generation artificial intelligence

**Authors**

Priyamvada Jadaun[1*†], Can Cui[1†], Sam Liu[1] and Jean Anne C. Incorvia[1*]

**Affiliations**

[1]Department of Electrical and Computer Engineering, The University of Texas at Austin, Austin, TX, 78712 USA.

*Corresponding author. Email: priyamvada@utexas.edu (PJ); incorvia@austin.utexas.edu (JACI). †These authors contributed equally to this work.

**Abstract**

Neuromorphic computing mimics the organizational principles of the brain in its quest to replicate the brain's intellectual abilities. An impressive ability of the brain is its adaptive intelligence, which allows the brain to regulate its functions "on the fly" to cope with myriad and ever-changing situations. In particular, the brain displays three adaptive and advanced intelligence abilities of context-awareness, cross frequency coupling and feature binding. To mimic these adaptive cognitive abilities, we design and simulate a novel, hardware-based adaptive oscillatory neuron using a lattice of magnetic skyrmions. Charge current fed to the neuron reconfigures the skyrmion lattice, thereby modulating the neuron's state, its dynamics and its transfer function "on the fly". This adaptive neuron is used to demonstrate the three cognitive abilities, of which context-awareness and cross-frequency coupling have not been previously realized in hardware neurons. Additionally, the neuron is used to construct an adaptive artificial neural network (ANN) and perform context-aware diagnosis of breast cancer. Simulations show that the adaptive ANN diagnoses cancer with higher accuracy while learning faster from smaller amounts of data and using a more compact and energy-efficient network than the state-of-the-art non-adaptive ANNs used for cancer diagnosis. The work further describes how hardware-based adaptive neurons can mitigate several critical challenges facing contemporary ANNs. Modern ANNs require large amounts of training data, energy and chip area and are highly task-specific; conversely, hardware-based ANNs built with adaptive neurons show faster learning from smaller datasets, compact architectures, energy-efficiency, fault-tolerance and can lead to the realization of general artificial intelligence.

**MAIN TEXT**

**Introduction**

Neuromorphic computing mimics the structural and functional principles of the human brain in hardware and aspires to replicate the brain's intellectual abilities [1]. In general, a remarkable and sought-after ability of the brain is its *adaptive* intelligence, which enables the brain to regulate its functions "on the fly" to effectively cope with ever-changing environments, situations, goals and rewards [2, 3, 4]. A specific instance of this adaptive intelligence is context-awareness, which is known to be the hallmark of cognition or high-level intelligence [1, 5]. Context-awareness is the ability of an intelligent agent to adapt its response to a certain situation or alter its decision in a scenario depending on underlying circumstances labeled 'context' [6]. It allows the brain to make complex, high-level decisions while considering multiple factors, to understand situations, react to new stimuli and make predictions from small datasets [7, 8]. Another adaptive cognitive function found in the brain is cross frequency coupling (CFC) which enables cognitive control of low-level neural signals by high-level knowledge [20]. One flavor of CFC is phase-amplitude CFC, where the phase of a global, low-frequency neural oscillation controls or adapts the amplitude of local, high-frequency neural oscillations, where neural oscillations are rhythmic activities in the brain that occur at all levels of neural organization [9]. CFC enables the brain to dynamically restructure its internal network, integrate various functional systems and control internal signaling [10, 11, 12, 13, 14]. A third adaptive ability called feature binding, enables the brain to combine different features of a perceived object into a coherent whole and integrate multi-modal information to build a coherent representation of the external world [15]. It is believed that in the brain, the phase of a global, low-frequency neural oscillation binds the features by controlling or adapting the frequency of local, high-frequency neural oscillations [16, 17, 18, 19, 20, 21]. Feature binding is known to be important for visual cognition [22] and is related to consciousness [23]. The realization of these cognitive abilities in neuromorphic computers can lead to artificial intelligence (AI) with transformative impact [24, 25].

The adaptive abilities of human intelligence described above originate in the adaptive ability of biological neurons along with the plasticity of biological synapses [26] mediated by collective dynamics of neural circuits. An adaptive neuron is one in which the neuron's properties, including its state, its transfer function and its state space, can be altered or modulated "on-the-fly" [27, 28]. A biological neuron is adaptive via neuromodulation, whereby a modulatory input to the neuron acts as a control signal and alters the neuronal properties to adapt its functioning to different situations [29, 30]. Through neuromodulation the brain can alter the amplitude and frequency of neural oscillations. This adaptive ability of biological neurons, specifically neuromodulation of neural oscillations, is critical for many additional cognitive processes, including information transfer, decision-making [31], memory [32], object representation [33, 34, 35], visual perception [36, 37] and attention [33, 38].

In addition to neuroscience, the importance of adaptive neurons is also well-recognized in software-based artificial neural networks (ANNs) and machine learning [27]. According to the universal approximation theorem [39], any continuous function can be represented by an ANN using conventional, non-adaptive neurons (labeled non-adaptive ANN in this work) to a degree of approximation while using one or more hidden layers and a finite number of weights. The approximation can be improved by increasing the number of hidden layers and the network size. In contrast, the Kolmogorov-Arnold representation theorem [40] implies that in an ANN built using adaptive neurons (labeled adaptive ANN in this work) where the non-linearity of the transfer function of every adaptive neuron can be selected, any continuous function can be represented exactly with an ANN using only one hidden layer. It is noteworthy that using adaptive neurons in an ANN allows a continuous function to be represented exactly (not just approximately) while using only one hidden layer. This implies that, in general, the expressive power of adaptive neurons in representing continuous functions supersedes that of non-adaptive neurons. In software-based ANNs, neurons with adaptive states and adaptive transfer functions have been implemented and shown to achieve superior classification performance with smaller ANN architectures than conventional non-adaptive neurons [27, 28, 41, 42, 43].

Despite the significant advantages of adaptive neurons, state-of-the-art hardware-based neurons are non-adaptive, such that these neurons have one input, a fixed transfer function and a given state space [44, 45, 46]. With the exception of spike-frequency adaptation shown in some hardware-based spiking neurons [47], most spiking neurons and all oscillatory neurons developed in hardware have a fixed state and transfer function. This is because neuromorphic implementation of neural oscillations has been very limited [48, 49] and neuromodulation of these neural oscillations has yet to be achieved. Amongst the three cognitive abilities mentioned above, context-awareness and CFC have not yet been realized in hardware-based neurons while feature binding has recently been realized in just a few hardware-based neurons [50].

Here, we design and simulate a previously unachieved, hardware-based adaptive neuron that incorporates both neuromodulation and neural oscillations. We utilize this neuron to realize the three cognitive abilities mentioned above, namely, context-awareness, CFC and feature binding. Additionally, the adaptive neuron is used to construct an adaptive ANN and perform context-aware diagnosis of breast cancer. Our simulations show that this adaptive ANN achieves diagnosis with higher accuracy while learning faster from smaller amounts of data and using a more compact and energy-efficient architecture than the state-of-the-art non-adaptive ANNs used for cancer diagnosis. This work further describes how hardware-based adaptive neurons are a fundamental improvement over conventional neurons and can mitigate several important challenges facing contemporary ANNs by enabling faster learning from smaller datasets, compact architectures, realization of general AI, energy-efficiency and fault-tolerance.

This adaptive neuron, named Tunable-SKyrmion-based Oscillating NEuron (T-SKONE), is designed using an artificial skyrmion lattice with five skyrmions hosted in a bilayer of insulating thulium iron garnet (TmIG) and platinum (Pt). The skyrmion lattice acts as a '*smart*' material that adapts its structure and properties in response to an external modulatory or control input. Simulations performed with MuMax3 [51] micromagnetic modeling show that when excited by an oscillating magnetic field, the neuron produces spin waves originating from skyrmion oscillations. The spin waves demonstrate a multi-frequency spectrum that results from skyrmion-skyrmion coupling [52, 53]. This neuronal spectrum consists of four distinct resonant modes, identified as counterclockwise gyration, breathing, and the hybridizations of both. The control input consists of an external current that re-arranges the locations of skyrmions in the lattice. This reconfiguration of the skyrmion lattice alters both the amplitude and frequency of the neuronal output, thereby altering the neuron's state, transfer function and state space, emulating neuromodulation.

T-SKONE is designed using an electrically reconfigurable lattice of skyrmions since spintronic nanodevices are highly attractive for neuromorphic computing due to their small footprint, high endurance and low power consumption [54, 55]. Moreover, skyrmions are particularly beneficial due to their small size and their ability to be created, manipulated, detected and erased by current or field [56]. However, the concept of leveraging smart materials to mimic neuromodulation is general and adaptive neurons could be realized with a variety of material systems including memristive, ferroelectric and magnetic domain wall materials [57, 58, 59, 60, 61, 62].

T-SKONE is used to realize context-awareness in hardware which is demonstrated with micromagnetic simulations. Here, the contextual information is fed into T-SKONE via the neuron's control input, such that changes in context alter the neuron's state, transfer function and its response. To demonstrate the potential of this context-aware ability of T-SKONE, neural network simulations are conducted to perform context-aware diagnosis of breast cancer. In these simulations, an adaptive ANN is constructed using adaptive neurons modeled after T-SKONE. Feature characteristics of breast mass biopsies are fed into the ANN as direct or conventional input and contextual patient medical data is fed into the ANN as control input. These simulations are benchmarked against state-of-the-art ANNs that diagnose cancer by fusing the biopsy feature data and contextual medical data without using adaptive neurons in a method known as multi-modal, multi-variate data fusion [63, 64]. The adaptive ANN is shown to outperform the non-adaptive ANN as it learns faster from smaller amounts of data and diagnoses cancer with higher final accuracy, while requiring a smaller architecture and less energy. Breast cancer was chosen as the task as it is one of the most common cancers in women, affecting 25% of all women with cancer worldwide [65] and early and accurate diagnosis of the disease enhances the survival rate [65].

The second cognitive ability shown by T-SKONE, CFC, is demonstrated by feeding a low-frequency wave into the control input of the neuron such that the phase of this wave modulates the amplitude of the high-frequency neuronal output. Taking inspiration from the brain where CFC is used to dynamically restructure the internal neural networks, we present the design of a structurally flexible, hardware-based ANN constructed from T-SKONEs. In this flexible ANN, CFC is utilized to dynamically switch the adaptive neurons 'on' or 'off', so as to reconfigure the network topology either on a pre-selected basis or at run-time. Flexible networks provide a general problem-solving tool that can be reconfigured for optimal solution of one task or across several tasks [66].

The third cognitive ability shown by T-SKONE, namely feature binding, is demonstrated by feeding a low-frequency wave into the control input of the adaptive neuron such that the phase of this wave modulates the frequency of the high-frequency neuronal output. A network of T-SKONES receives visual information about two objects, a 'red circle' and a 'blue cross'. Two different features (shape and color) of the same object are encoded by the neurons at the same frequency such that they can be correctly bound together via synchronization to construct a coherent percept of the original object.

Therefore, hardware-based adaptive neurons can realize previously unachieved cognitive capabilities with far-reaching applications. Context-aware computation demonstrated here can have sweeping impact on human-machine collaboration, personalized healthcare, advanced manufacturing and education [25, 67, 68, 69, 70]. Similarly, CFC-based flexible ANNs presented here can achieve general, energy-efficient and compact AI which upon integration with 5G and IoT technologies would lead to smart agents on the edge impacting numerous facets of everyday life [71]. Moreover, feature binding enables multi-modal information fusion which can transform autonomous vehicles, neuro-prosthesis, wearable health technology, agriculture and climate control [72, 73, 74, 75].

More generally, hardware-based adaptive neurons can mitigate several critical challenges facing contemporary ANNs. Modern ANNs demand huge volume of training data [76, 77], unsustainably large energy and chip area when implemented on CMOS circuits [78] and are highly specific or non-general in their applications [24, 79]. In contrast, adaptive neurons like T-SKONE can mitigate these challenges by enabling ANNs that learn faster from smaller datasets, have compact architectures, are energy-efficient, are fault-tolerant and realize general artificial intelligence.

Results

**Design of hardware-based adaptive oscillatory neuron** - A schematic of T-SKONE is shown in Fig. 1 (a). The skyrmion lattice comprises five nanotracks, (with a nanotrack along $y$), each hosting a single skyrmion (blue). The five skyrmions are labeled *Sk1 - Sk5* along increasing $x$, as shown in Fig. 1 (a). The neuron is designed to mimic neuromodulation (see Fig. 1 (b)). To precisely arrange the skyrmions in the lattice and prevent skyrmion dislocation upon application of an oscillatory magnetic field, reduced perpendicular magnetic anisotropy (PMA) regions are used as skyrmion pinning sites (shadowed regions in Fig. 1 (c)). Such reduced PMA regions can be created using ion implantation [80]. Since the skyrmion boundary with in-plane magnetization tends to be located in the low-PMA regions to reduce the overall magnetostatic energy [81], the skyrmions can be softly pinned into lattice sites A or B (Fig. 1 (c)). The neuron receives two types of inputs: (i) direct or driving input in the form of an excitatory magnetic field input ($H_{RF}$) along $x$ applied to all skyrmions through a gold strip line antenna, which induces localized skyrmion oscillations and produces spin waves [82, 83], and (ii) control or modulatory input in the form of charge currents ($J_1 - J_5$) along $y$ applied to individual skyrmions through the spin Hall metal Pt overlayers (purple), which independently move the skyrmions along the nanotracks and position them in site A or B. As the skyrmion lattice configuration is modified (e.g. between Configurations I and II as shown in Fig. 1 (c)), so are the skyrmion coupling strengths and therefore their resonant frequencies and amplitudes of oscillation. Thus, the response of the neuron to the direct input can be modulated by the control input, mimicking neuromodulation. The outputs from the skyrmion nanotracks can be read from inductive antennas placed below individual nanotracks in the form of AC voltages or carried by coplanar waveguides composed of an insulating magnet (YIG) in the form of spin waves.

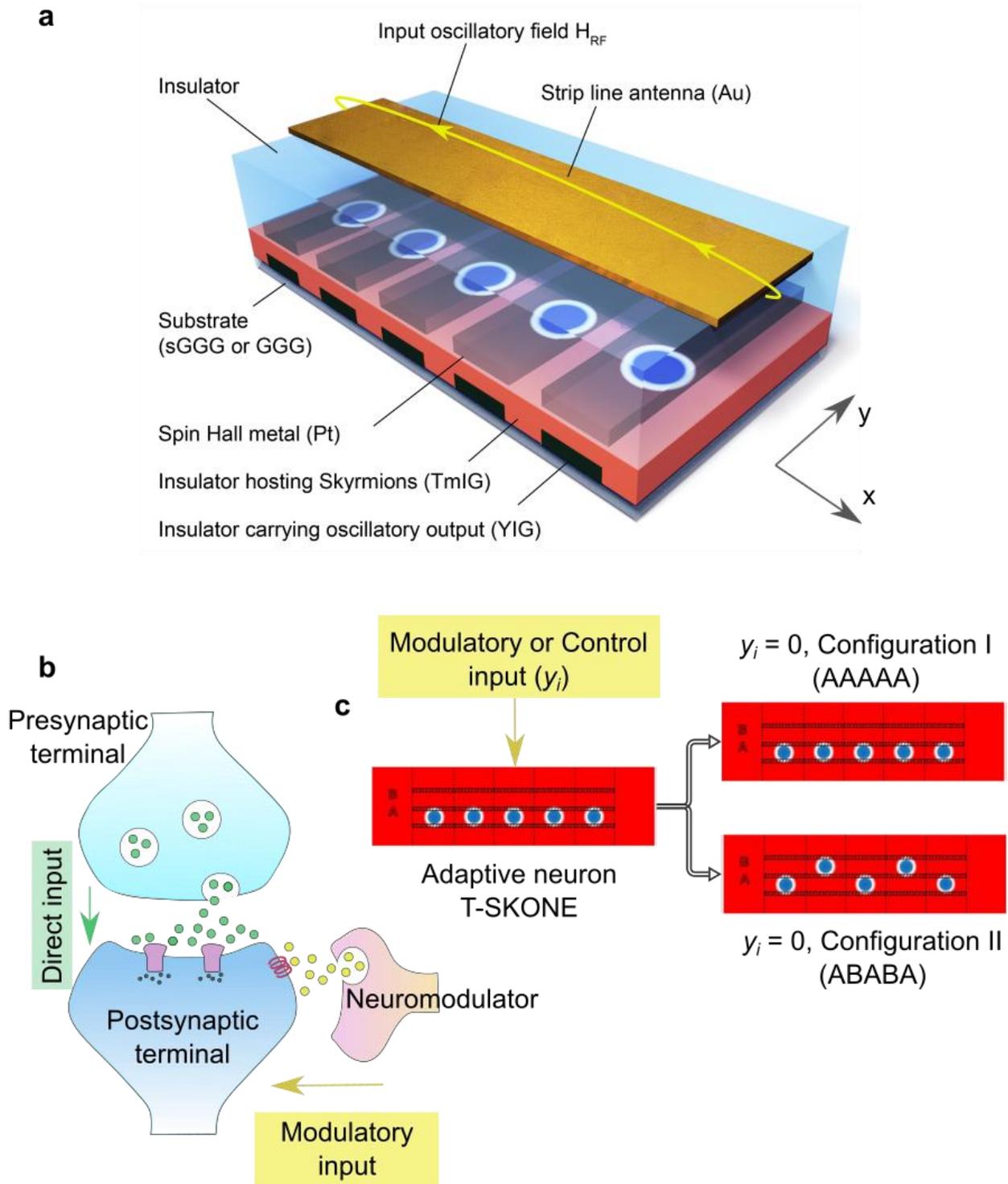

FIG. 1. Schematic of the adaptive oscillatory neuron. (a) Structure of the neuron. TmIG (salmon): insulating magnet hosting the skyrmion lattice. $H_{RF}$: excitatory magnetic field input carried by the strip line antenna. $J_1 - J_5$: modulatory current inputs carried by the spin Hall metal (Pt). YIG: spin wave output readout. (b) Schematic for neuromodulation in the brain. Direct input is carried from the presynaptic terminal to the postsynaptic terminal via neurotransmitters (green circles) and modulatory input (yellow circles) is carried from the neuromodulator to the postsynaptic terminal. (c) Schematic for neuromodulation in the skyrmion lattice. The control input ($y_i$) reconfigures the skyrmion lattice, with skyrmions (blue circles) arranged in Configuration I (AAAAA, top) for $y_i=0$, and II (ABABA, bottom) for $y_i=1$. The shadowed regions have reduced perpendicular magnetic anisotropy, designed to softly pin the skyrmions.

Recent work has reported the observation of above and near room-temperature skyrmions in a bilayer heterostructure composed of a magnetic insulator, thulium iron garnet (TmIG, $Tm_3Fe_5O_{12}$) in contact with a spin Hall metal (Pt) [84]. There have also been experimental observations of domain wall motion in TmIG/Pt bilayers driven by spin orbit torque [85] [86]. The TmIG/Pt system provides innate electrical isolation to the individual nanotracks, while at the same time allowing the skyrmions to couple magnetically. However, since the skyrmion phase in TmIG/Pt typically exists at and above 360 K, more work is required for the design of room-temperature stable insulating skyrmion materials.

**Simulation of the adaptive neuronal function** - The oscillatory dynamics of T-SKONE was simulated for two lattice configurations, Configuration I (AAAAA) and II (ABABA), using the MuMax3 solver [51], as described in Methods. The fast Fourier transform (FFT) power density spectra of the skyrmion core trajectories was calculated (see Methods) and is shown in Fig. 2 (a-b), with Configuration I in solid blue and Configuration II in dashed magenta. As the configuration of the neuron changes, the peaks of the spectrum that signify the resonant frequencies also shift, which demonstrates that the neuron is capable of Frequency Modulation. Figure 2 (a) shows two resonant peaks for both configurations at around 0.80 GHz and 1.04 GHz corresponding to Mode 1 and Mode 2, respectively. The FFT spectra when zoomed-in and extended up to 2.0 GHz is shown in Fig. 2 (b), where two more resonant peaks around 1.22 GHz (Mode 3) and around 1.85 GHz (Mode 4) can be seen. In total, there are $2^5$ neuronal configurations available in this design and more work is required to determine the configurations with notably distinct dynamics. T-SKONE is estimated to utilize ultra-low power of < 0.1 pJ/oscillation at 1 GHz and is ultra-compact at 1200 nm length compared to a CMOS neuron (265 pJ/oscillation, > 30 μm) and a spin-torque oscillator (3 pJ/oscillation, 300 nm) [87], while being significantly more computationally powerful. The power utilized by T-SONE is estimated based on the driving current in the strip line antenna required to produce an external AC magnetic field of $10\ Oe$ and a DC magnetic field of $240\ Oe$, with the antenna assumed to be 300 nm above the neuron.

To demonstrate Amplitude Modulation, the neuron was excited by a sinusoidal magnetic field as described in Methods. Figure 2 (c-f) show the amplitudes of oscillations of the skyrmion cores along *x* at driving frequency ranges centered on (c) Mode 1, (d) Mode 2, (e) Mode 3, and (f) Mode 4. There are multiple instances of a clear difference in the amplitudes of oscillations at a given frequency for Configurations I (solid blue) and II (dashed magenta), which demonstrates Amplitude Modulation. For instance, at 0.80 GHz (Mode 1) *Sk3* has a larger amplitude of oscillation for Configuration II than I, while *Sk5* shows a larger amplitude for Configuration I than II, both of which are consistent with the FFT power spectrum. It is worth noting that the small driving field ($10\ Oe$) required to induce large oscillation amplitudes (up to 8 nm) is highly promising for ultra-low power operation of the neuron. Amplitude Modulation along *y* is shown in Supplementary Information S1.

We now discuss the four resonant modes identified above in more detail. Figure 3 shows the oscillatory response of the neuron while in Configuration I, upon excitation with a sinusoidal magnetic field. The first and second columns of Fig. 3 (a-d) plot the oscillations for every skyrmion core along $x$ and $y$, respectively, for resonant modes (1-4). The skyrmion cores oscillate sinusoidally with frequencies locked to the driving input and with a wide range of amplitudes. To reveal the physical origin of these modes, topological charge density maps of the T-SKONE are plotted at times $t = 0, T/4, T/2$ and $3T/4$, where $T$ is the respective time period of oscillation. These maps are shown in the four columns on the right of Fig. 3 and demonstrate that every resonant mode originates from a distinct type of skyrmion dynamics. In general, the oscillation amplitudes are larger for the outer skyrmions (*Sk1* and *Sk5*) than amplitudes for inner skyrmions (*Sk2 - Sk4*). This is likely because the dominant form of interaction is the skyrmion-skyrmion repulsion which suppresses the oscillations of the inner skyrmions.

Figure 3 (a) shows the skyrmion oscillation for Configuration I at Mode 1 (0.80 GHz). The topological charge density plots demonstrate that this is a breathing mode, in which the skyrmion cores periodically expand and contract in size. At this frequency the skyrmions are out-of-phase with their neighbors, which is favored by the skyrmion-skyrmion repulsion. Figure 3 (b) plots the output for Configuration I at Mode 2 (1.04 GHz). From the topological charge density maps, a mixture of the two oscillations instead of complete cycles of breathing mode or gyration mode are observed over the time period of oscillation, which indicates that this mode is a hybridization between these two pure modes. The amplitudes for the inner skyrmions (*Sk2 - Sk4*) are comparable, and their oscillations while not perfectly phase-locked are similar in their phase ordering. This is expected due to the hybridized nature of Mode 2.

Skyrmion oscillations at Mode 3 (1.22 GHz) are shown in Fig. 3 (c). Topological charge density plots reveal the origin of this mode to be a CCW gyration mode with all the skyrmions synchronized at 1.22 GHz. Finally, Fig. 3 (d) demonstrates oscillatory dynamics at Mode 4 (1.85 GHz) where oscillations for all skyrmions are generally small (< 1 nm). Topological charge density maps reveal this to be a highly complex mode originating from the hybridization of breathing, counterclockwise and clockwise gyration.

With few exceptions, the largest oscillatory amplitudes were seen for Modes 1 and 3 and the smallest were seen for Mode 4. This is likely because Modes 1 and 3 are pure modes whereas Modes 2 and 4 are hybridizations of the pure modes. The frequency range corresponding to Mode 2 lies between that of the pure Modes 1 and 3, since Mode 2 is a hybridization of Modes 1 and 3. The discovery of hybridized modes in skyrmion lattices brings a particularly rich and coupled dynamics to the oscillatory neuron, enhancing its utility for neuromorphic applications. When the neuron in Configuration II was excited with the same frequencies as above, it also demonstrated four modes with the same physical origin, but with different oscillation amplitudes (see Supplementary Information S2). Additionally, Amplitude Modulation and Frequency Modulation were demonstrated using the metallic CoFeB/heavy metal bilayer system (see Supplementary Information S3).

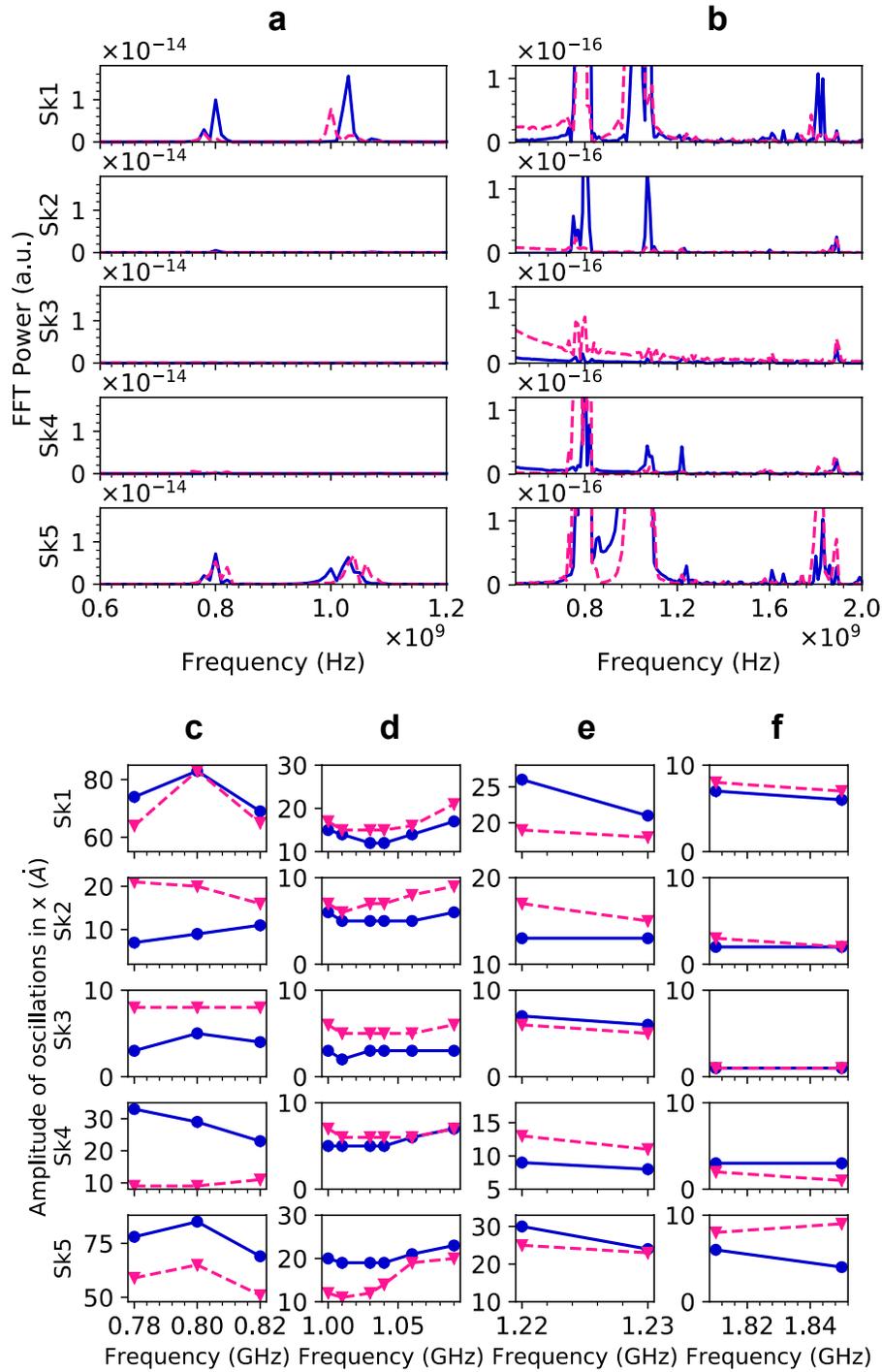

FIG. 2. Coupled dynamics of T-SKONE. (a) Frequency Modulation: Fast Fourier transform (FFT) power density spectra of the neuron for Configuration I (solid blue line) and Configuration II (dashed magenta line), for all five skyrmions (*Sk1 - Sk5*). (b) Zoomed-in plot of (a). (c-f) Amplitude modulation: Amplitudes of oscillation for all five skyrmions shown at a range of frequencies for Configuration I (blue circles) and Configuration II (magenta triangles).

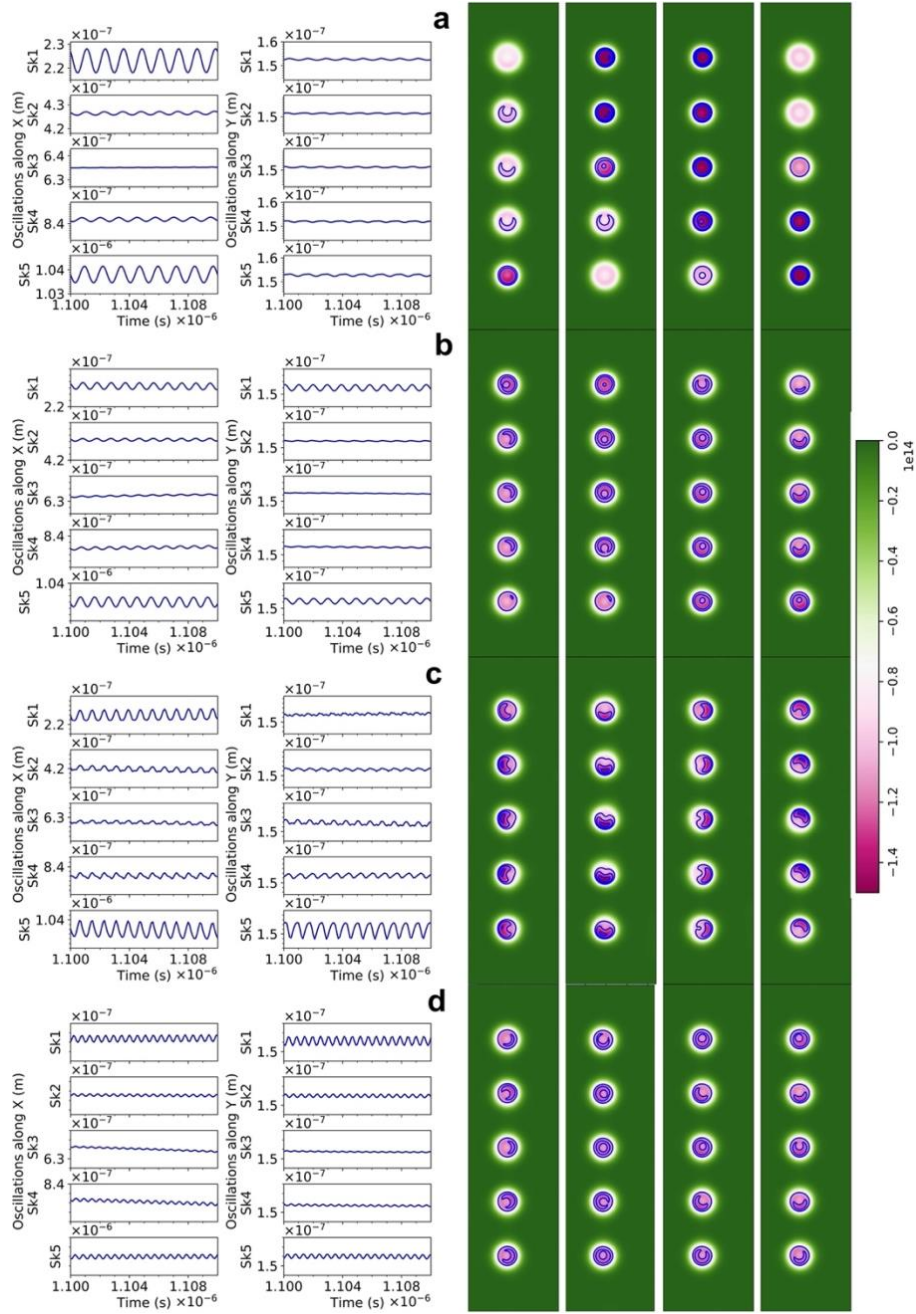

FIG. 3. Resonant modes of T-SKONE in Configuration I when excited by a sinusoidal driving field of frequency (a) 0.80 GHz (Mode 1), (b) 1.04 GHz (Mode 2), (c) 1.22 GHz (Mode 3) and (d) 1.85 GHz (Mode 4). The two left most columns plot the output oscillations along *x* and *y*, respectively. The four columns on the right plot the topological charge densities at times $t = 0, T/4, T/2, 3T/4$, where $T$ is the respective time period of oscillation. The color maps for topological charge density ranges from $-1.5 \times 10^{14} m^{-2}$ (magenta) to zero (green).

**Design of an ANN using T-SKONE** – The input-output characteristics of T-SKONE obtained using micromagnetic simulations are now mapped into a model adaptive neuron and an ANN is constructed using this neuron, as shown in Fig. 4. Since T-SKONE has many possible configurations and its output can be selected from any of the five skyrmion nanotracks, the exact neuronal outputs and the network architecture will depend on the specific task at hand. However, this section presents a general method used to construct an ANN using the adaptive neurons.

Here, we consider a fully connected, feed-forward ANN with 2 layers. Neurons in the first layer are in an adaptive state and receive 2 inputs, a conventional or direct input ($x_i$) and a control or modulatory input ($y_i$). The direct input ($x_i$) is fed to T-SKONE ($T_i$) encoded in the amplitude of a sinusoidal magnetic field ($H_{RF}$) with a frequency of 0.80 GHz. This field excites the breathing mode of the skyrmion lattice in T-SKONE. The control input ($y_i$) is fed to T-SKONE encoded in modulatory currents ($J_1 - J_5$) that reconfigure T-SKONE and alter its state, as shown in Fig. 4 (a). Control input $y_i = 0, 1$ corresponds to Configuration I, II, respectively.

Here, the output of T-SKONE is designed to be detected by an inductive antenna placed below the nanotrack for skyrmion 4 (fourth skyrmion along $x$ in Fig. 1 (a)). This induced AC voltage is proportional to the average change in magnetization in the nanotrack $\Delta M$ ($V_{peak} \approx \mu_0 \Delta M A \varpi$) [88], where $\mu_0$ is the vacuum permeability, $A$ is area of the antenna and $\varpi$ is the oscillation frequency. Therefore, $\Delta M/M_s$ is used as the effective T-SKONE output in this task, where $M_s$ is the saturation magnetization. This effective output was calculated using micromagnetic simulations by averaging the change in magnetization over four distinct regions of the nanotrack, i.e., the skyrmion core, the skyrmion boundary, the region between the core and boundary and the region of the nanotrack outside the skyrmion. The neuronal output ($\Delta M/M_s$) plotted in Fig. 4 (b) in green, is notably different for different control inputs ($y_i = 0,1$). Therefore, the control input modulates the state and transfer function of the adaptive neuron. The neuron's input-output characteristics were fitted to a polynomial transfer function (plotted in Fig. 4 (b) in gray) to obtain a model transfer function. The transfer function $f(x_i, y_i)$ for i$^{th}$ neuron is given by:

$$f(x_i, y_i) = \begin{cases} -0.00038 x_i^2 + 0.04413 x_i + 0.00300, & for\ y_i = 0 \\ -0.00049 x_i^2 + 0.05347 x_i - 0.42914, & for\ y_i = 1 \end{cases}$$

The outputs of the neurons in the first layer $f(x_i, y_i)$ are sent through synaptic weights $w_{ij}$, added and sent as direct input to neurons in the second layer, as shown in Fig. 4 (c).

In this general example, the neurons in the second layer are T-SKONEs (skyrmion arrays) kept in a fixed state, namely, Configuration I. The input received by the j$^{th}$ neuron of the second layer is $u_j = \sum_{i=1}^{q} w_{ij} f(x_i, y_i)$. The transfer function $g(u_j)$ of the j$^{th}$ neuron of the second layer is obtained by mapping the input-output characteristics of T-SKONE in Configuration I to a polynomial function and is given by:

$$g(u_j) = -0.00347 u_j^3 + 0.34476 u_j$$

*Adaptive cognition implemented with...*     11     *Jadaun et al.*

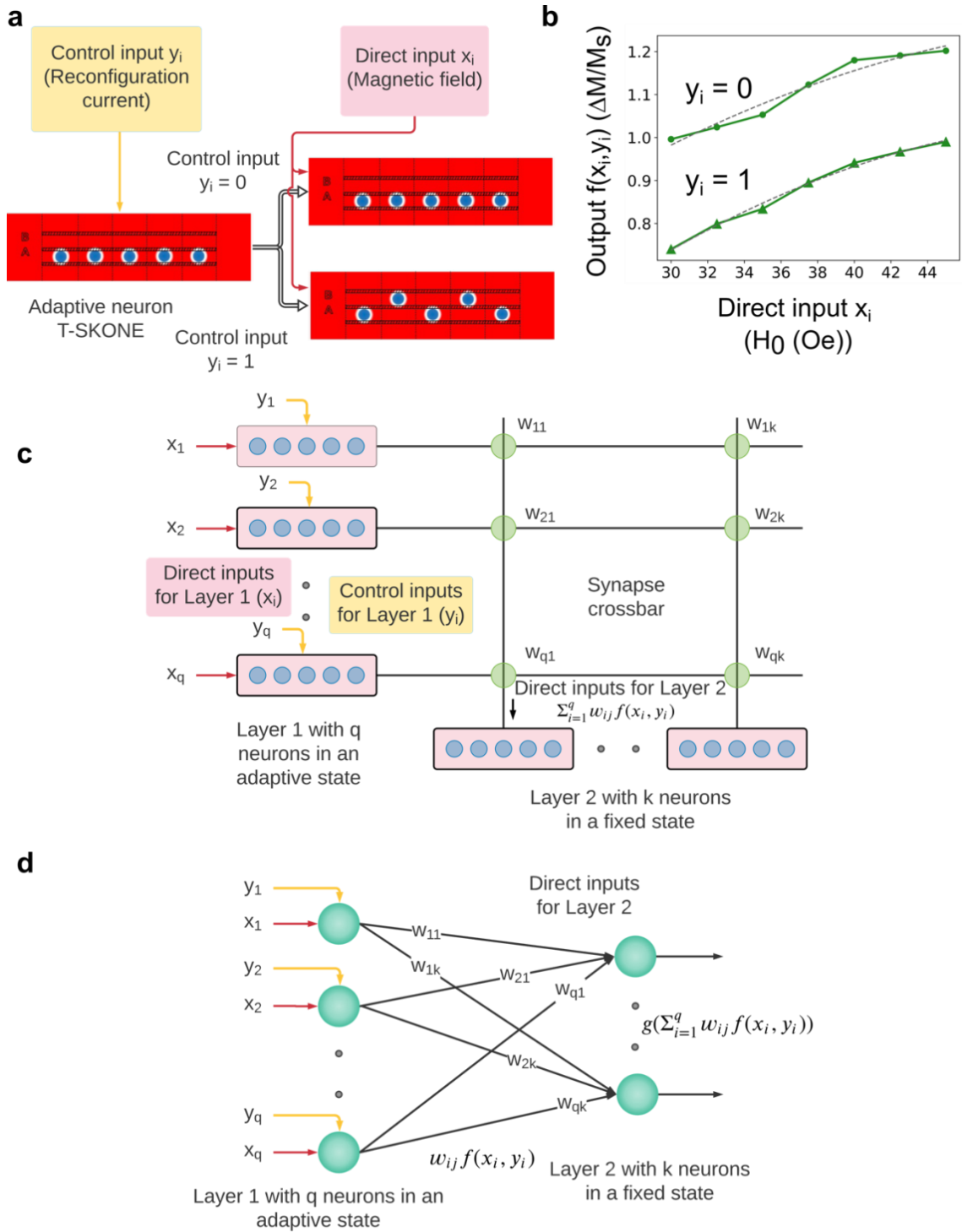

Fig. 4: Mapping of skyrmion arrays (T-SKONEs) into an adaptive ANN. (a) Schematic for adaptive behaviour in a single T-SKONE. The neuron receives a control input $y_i$ that reconfigures the neuronal state (b) Transfer function $f(x_i, y_i)$ for the adaptive neuron T-SKONE which is modified by the control input $y_i$. The transfer function is calculated using micromagnetic simulations (green) and converted into an analytical model (gray) that is used for neural network simulations. (c) Diagram for a hardware ANN using T-SKONEs. In layer 1, T-SKONEs receive two inputs, direct ($x_i$) and control ($y_i$) with the latter modulating the neuronal state. The outputs from layer 1 are fed into layer 2 via spintronic synapses and T-SKONEs in layer 2 are kept in a fixed state. (d) The equivalent network model for the ANN shown in (c).

We note that $g(u_j)$ is not equal to $f(x_i, y_i = 0)$, even though in both cases T-SKONE is in Configuration I. This is because the range of direct inputs for layer 1 ($x_i$) and layer 2 ($u_j$) are different. In the first layer, $x_i$ is restricted to values between 30 and 45 $Oe$ to prevent any overlap between the output values seen for the two configurations, while in the second layer $u_i$ can take any value between 0 and 45 $Oe$. Therefore, $g(u_j)$ and $f(x_i, y_i = 0)$ are fitted to functions with different ranges. The final output of the j$^{th}$ neuron of the second layer is $g(\sum_{i=1}^{q} w_{ij} f(x_i, y_i))$.

In a hardware implementation of this ANN, using $\Delta M \sim M_s$ (saturation magnetization) gives an estimated AC output voltage for T-SKONE in the range of 0.01 mV. This AC output can be amplified and fed to a crossbar of domain wall-based synapses that implement the stochastic gradient descent learning algorithm [89] and apply weights $w_{ij}$. The output AC voltages from this synaptic layer can be fed as direct inputs to the second neural layer in the ANN. The need for amplification, which hampers the energy efficiency of this network, can be mitigated by using skyrmion nanotracks with larger $M_s$ or designing spin wave-based synapses that directly connect subsequent T-SKONE layers without using inductive antennas.

**Demonstration of context-aware diagnosis of breast cancer using T-SKONE** – Context-awareness is a key ability of the brain, where context refers to information that characterizes the situation of an entity, e.g., place, time, emotional state, etc [90]. In this work, context-aware tasks are defined as tasks where classification or prediction is performed using (i) direct information which is defined as the data that is being classified and (ii) contextual information which is defined as background data that is related to the direct information but is not itself being classified. As an example, here we implement context-aware diagnosis of breast cancer where biopsy image features of breast mass are classified to predict whether the biopsy results are malignant or benign and patient's medical data provides important background information to help make the diagnosis. Therefore, the biopsy image features that are being classified are direct information, and the patient's medical data is contextual information. Context-awareness includes a large class of applications encompassing brain-machine interfaces, biomedicine, IoT, robotics etc.

Due to the absence of a well-known, multi-modal cancer dataset containing information about both biopsy images and patient medical data, we constructed such a dataset by appending the Breast Cancer Wisconsin (Diagnostic) dataset (comprising biopsy image features) [91] with information regarding patient medical data (comprising attributes *$\alpha_i$*). The four selected attributes selected were: (i) patient is female, (ii) patient is older than 50 years, (iii) patient has a Body Mass Index (BMI) in the obese range, and (iv) patient has a large intake of alcohol. The modified dataset was constructed to represent statistical correlations seen in reality, details about which are described in Methods. Breast cancer was selected as the task since it can be implemented as a context-aware problem unlike more commonly used benchmarking approaches like the MNIST image recognition task. In addition, breast cancer is selected since it is the most commonly diagnosed cancer in women, affecting 2.1 million women each year and is the leading cause of female cancer deaths. Early and accurate detection can enhance survival rates among breast cancer patients [92].

In general, context-aware tasks can be implemented using adaptive neurons like T-SKONE by encoding the contextual information in the neuron's control input. Changes in context, alter the control input which in turn alters the neuron's state, transfer function and therefore the neuronal response, implementing context-awareness. In this specific task, an adaptive ANN was constructed from T-SKONEs in the same manner as the general method described above (see Fig. 4). The ANN comprised a fully connected, feed-forward network with 2 layers. Feature characteristics of biopsies were fed to T-SKONE ($T_i$) in the first layer as direct input ($x_i$) encoded in the amplitude of a sinusoidal magnetic field ($H_{RF}$) with a frequency of 0.80 GHz and a range of values between 30 and 45 $Oe$. Patient medical data was fed to T-SKONE ($T_i$) in the first layer as control input ($y_i$), such that $\alpha_i$ = 'F' or 'T' corresponded to $y_i = 0$ or 1, respectively.

The state space of the $i^{th}$ T-SKONE in the first layer ($T_i$) is given by:

$$\begin{cases} [0.996, 1.202]\ \Delta M/M_s, & for\ y_i = 0 \\ [0.740, 0.990]\ \Delta M/M_s, & for\ y_i = 1 \end{cases}$$

As before, T-SKONE was mapped onto a model neuron by fitting the former's input-output characteristics to a polynomial transfer function which is plotted in Fig. 5 (a) and is given by:

$$f(x_i, y_i) = \begin{cases} -0.00038 x_i^2 + 0.04413 x_i + 0.00300, & for\ y_i = 0 \\ -0.00049 x_i^2 + 0.05347 x_i - 0.42914, & for\ y_i = 1 \end{cases}$$

In this simulation, the output from Layer 1 was fed to a series of software synapses that were treated as perfect double-precision weights $w_{ij}$. The synapses were trained by applying the Adam optimization algorithm [93], a form of adaptive momentum stochastic gradient descent. The synaptic output was fed as direct input to neurons in the second layer which were modeled after T-SKONES kept in a fixed state (Configuration I). The input into a $j^{th}$ neuron was $u_j = \sum_{i=1}^{q} w_{ij} f_i(x_i, y_i)$ and the range of this input was kept between 0 and 45 $Oe$ by scaling synaptic weights. Neurons in the second layer had a state space of [0, 1.202] $\Delta M/M_s$ and a transfer function $g(u_j) = -0.00347 u_j^3 + 0.34476 u_j$, which is shown in Fig. 5 (b).

In the first simulation, the topology of the network was chosen such that Layer 1 had 30 neurons in the adaptive state and Layer 2 had 2 neurons in the fixed state, see Fig. 5 (c). This is because each biopsy image sample in the breast cancer dataset had 30 attributes and the samples were classified as either malignant or benign. The contextual inputs ($\alpha_1, \alpha_2, \alpha_3, \alpha_4$) were fed into 7, 7, 8, and 8 neurons of Layer 1, respectively. The neural network simulation was implemented in PyTorch [94]. 20% of the samples were separated from the training set and used for validation. The network training was applied for 800 epochs. Results from this simulation are plotted in Fig. 5 (d)-(f) in dark green.

To benchmark the performance of the adaptive ANN constructed above, we also carried out cancer diagnosis with an ANN built from non-adaptive neurons. State-of-the-art in machine learning for context-aware cancer diagnostics uses multi-modal data fusion [63, 64, 95], where both biopsy image features and patient medical data are fed as direct input to conventional, non-adaptive neurons. To implement this method, we constructed an ANN that uses T-SKONEs kept in a fixed state (Configuration I). The biopsy data was encoded in the direct input ($x_i$) as before, and the medical data was also encoded in the direct input such that $\alpha_i$ = 'F' or 'T' corresponded to $x_i = 0$ or 1, respectively. Layer 1 of the network consisted of a total of 34 neurons, with 30 neurons that processed the biopsy feature data and had a state space of [0.996, 1.202] $\Delta M/M_s$ and 4 neurons that processed medical data and had a state space of {0.996, 1.202} $\Delta M/M_s$. The $i^{th}$ neuron of Layer 1 had the transfer function $f'(x_i) = -0.00038x^2 + 0.04413x + 0.00300 = f(x_i, y_i = 0)$. Layer 2 was kept the same as for the adaptive ANN (see Fig. S5 in Supplementary Information). We label this simulation 'direct data fusion' and plot its results in Fig. 5 (d) & (f) in orange.

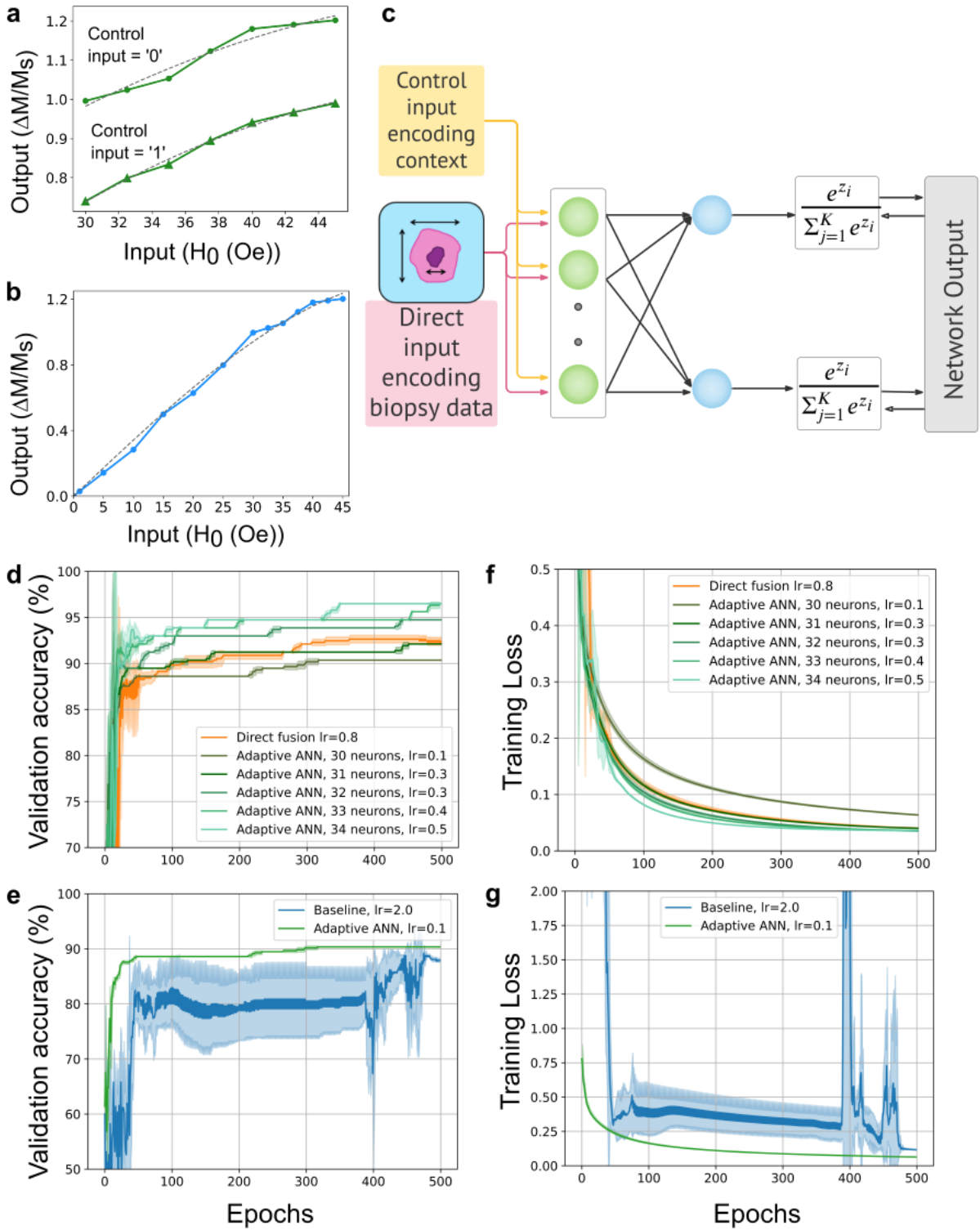

FIG. 5. Context-aware diagnosis of breast cancer using T-SKONE. (a) Transfer function for T-SKONE used in Layer 1. The control input ($y_i$) modulates the configuration of the skyrmion array and alters the neuron's transfer function. (b) Transfer function for T-SKONE in fixed state used in Layer 2. (c) Diagram of adaptive ANN used for diagnosis, with neurons in Layer 1 (blue circles) receiving biopsy image data (direct input) and patient medical data (control input). (d-e) Validation accuracy and (f-g) Loss of information for adaptive ANN (green), direct data fusion (orange) and baseline ANN (blue). The adaptive ANN consistently learns faster from smaller amounts of data and achieves higher accuracy while using a smaller network.

An additional benchmarking simulation was performed where breast cancer was diagnosed using only the baseline data of biopsy image features. The ANN was constructed exactly as the benchmark described above, except that Layer 1 only had 30 neurons and only took in biopsy data encoded in the direct input ($x_i$) (see Fig. S5 in Supplementary Information). We label this simulation 'baseline' and plot its results in Fig. 5 (e) & (g) in blue.

To verify the performance of the adaptive ANN with changing network topologies and for a fair comparison with the two benchmarking simulations that each have 34 and 30 neurons in Layer 1, we varied the number of neurons in the Layer 1 of the adaptive ANN between 30 and 34. This was done by adding neurons to the layer that only took in a constant input. For robust benchmarking, the learning rate was individually optimized for every simulation.

A consistent feature of the ANN built from adaptive neurons is its ability to learn faster from smaller amounts of data than ANNs built from non-adaptive neurons, which persists even with changes in network topology. This can be seen from the smaller optimized learning rate ($lr$) for the adaptive ANNs ($0.1 \leq lr \leq 0.5$) than for the non-adaptive ANNs used in direct data fusion ($lr$=0.8) and baseline ($lr$=2.0) simulations. Additionally, the validation accuracy in the initial epochs is higher (see Fig. 5 (d) & (e)) and training loss is lower (see Fig. 5 (f) & (g)) for adaptive ANNs (shown in green) than for the non-adaptive ANNs, signifying faster learning by the adaptive neurons.

While the adaptive ANN consistently learns faster than the benchmark ANNs, the final accuracy of classification depends on the network topology as well. As Fig. 5 (d) shows, the final validation accuracy of the adaptive ANN increases with increase in the number of neurons in Layer 1. As can be seen, the adaptive ANN with 32-34 neurons in Layer 1 classifies with higher accuracy than the ANN for direct data fusion which has 34 neurons in Layer 1 (see Fig. 5 (d)). Hence, the adaptive ANN reaches higher classification accuracy while using a smaller architecture. The adaptive ANN with 30 neurons in Layer 1 also classifies more accurately than the baseline ANN with 30 neurons. The accuracy and loss results were averaged across 5 randomized weight initializations. The shaded regions of Fig. 5 (d)-(g) show the standard deviation of the relevant metric at a particular epoch.

**Fundamental benefits of T-SKONE-based context-aware ANNs**- Hardware-based adaptive neurons like T-SKONE are a fundamental advancement over non-adaptive neurons and can mitigate several challenges facing conventional ANNs that are built using non-adaptive neurons.

*1. Faster learning using smaller datasets* – A severe challenge facing conventional ANNs is their requirement of huge volumes of training data [76, 77]. In contrast, our results show that an ANN with adaptive neurons consistently learns faster from smaller amounts of data than conventional ANNs in context-aware tasks. Moreover, in software-based ANNs, adaptive neurons have been verified to learn faster than non-adaptive neurons in a wide variety of tasks using various numerical approaches [27, 28, 41, 42, 43].

*2. Energy-efficiency and compact architecture* – Another significant challenge is that software-based ANNs implemented with conventional CMOS circuits use unsustainably large energy and area [78]. In contrast, T-SKONE uses ultra-low power and is ultra-compact. Additionally, it is well known that neurons with adaptive transfer functions can express a desired function using more compact architectures than non-adaptive neurons [40]. Therefore, T-SKONE-based ANNs will use less energy and chip area than software-based ANNs and many hardware-based non-adaptive ANNs.

*3. Realizing general AI* – Conventional ANNs are adept at highly specific tasks and are not proficient outside that domain [24]. For instance, conventional ANNs cannot effectively deal with changing environments, contexts, goals or rewards their applications [24, 79]. This lack of 'general intelligence' is a core challenge facing modern ANNs [24]. In this work, contextual information controls the functioning of adaptive neurons enabling the ANN to deal with dynamically varying context more effectively than conventional ANNs. Similarly, environmental signals, goals and rewards can also be used to control the functioning of T-SKONE, to realize ANNs adept at performing dynamic tasks. An example of this is demonstrated for a human-machine interaction task (See Supplementary Information S6). Here, the neuron receives a direct input representing a human spoken command and a control input representing environmental safety, specifically whether a box is safe to open. Upon receiving a 'go' command from the human being, the neuron can adaptively decide whether or not to open the box, depending on whether the box is safe or not. These results present a roadmap for realizing general AI using T-SKONE-based ANNs.

Additionally, a recent breakthrough in the development of general AI involves Neuro-symbolic AI. A combination of Deep Neural Networks (DNNs) and Symbolic AI, Neuro-symbolic AI creates symbolic models that capture the underlying relationships between factors and forces, are augmented with contextual information, and can predict situations that are yet to be encountered [96, 97]. As T-SKONE-based ANNs are fundamentally superior at processing contextual information than conventional ANNs, the combination of adaptive ANNs with Symbolic AI could be highly attractive for realizing general AI. Furthermore, an important concept in the search for general AI is top-down knowledge, where ''understanding'' is encapsulated in feedback signals from high-level to low-level networks [98]. The control input of T-SKONE is innately well-suited to carrying this top-down knowledge enabling high-level networks to modulate the functioning of low-level networks.

*4. Fault-tolerance* - An additional benefit of adaptive neurons is that at any time, different neurons in the ANN are in different states and use different transfer functions. Thus, the errors produced by these neurons are less correlated than in conventional ANNs, which leads to greater fault-tolerance [99].

**Demonstration of cross frequency coupling (CFC) and structurally flexible ANNs using T-SKONE** – Cross frequency coupling (CFC) is a cognitive mechanism in which a global, low-frequency neural oscillation *controls or modulates* local, high-frequency neural oscillations in the brain (see Fig. 6 (a)). Involved in 3 cognitive operations: (i) multi-item representation, (ii) long-distance communication, and (iii) stimulus parsing [21], CFC enables the brain to alter the architecture of its NNs, coordinate message transfer between different NNs and integrate information across several time scales [32]. Here, CFC is realized by feeding a low-frequency wave into the control input of T-SKONE (see Fig. 6 (b)) such that the phase of this wave modulates the amplitude of the high-frequency neuronal output.

CFC is demonstrated using micromagnetic simulations for two cases. In both cases, the control input was a low-frequency modulatory wave of 10 MHz. The direct input fed to T-SKONE was a high-frequency sinusoidal magnetic field with frequency 0.80 GHz in case 1 and 1.85 GHz in case 2. The neuron's output was taken as the amplitude of oscillation of *Sk2* along *x* for case 1, and that of *Sk5* along *x* for case 2. Here, the amplitude of skyrmion core oscillation is taken as the effective output since it is easy to calculate while being related to the change in magnetization in the nanotrack ($\Delta M$), the latter being the effective output taken previously. Reconfiguration was carried out by applying an electric current pulse of density $10^{11}$ A/m$^2$ and duration of 0.8 ns using a spin Hall angle of 0.02 corresponding to Pt. Figure 6 (c) and (d) plot the output oscillations of *Sk2* and *Sk5* along *x*, respectively, and clearly demonstrate that the low-frequency modulatory input controls the amplitude of the high-frequency neuronal output, implementing CFC.

Taking inspiration from the brain, we present the design of a structurally flexible ANN constructed from T-SKONEs that utilizes CFC to dynamically alter the network architecture. Structurally flexible networks have been previously proposed in software-based ANNs using the techniques of growing and pruning [100]. Hardware-based flexible ANNs have achieved structural flexibility by incorporating switches into the network circuit that can switch neurons on or off to alter the network topology [66]. Here, we utilize CFC to dynamically turn the adaptive neuron on or off via a control signal without the need for extra switches in the network circuit.

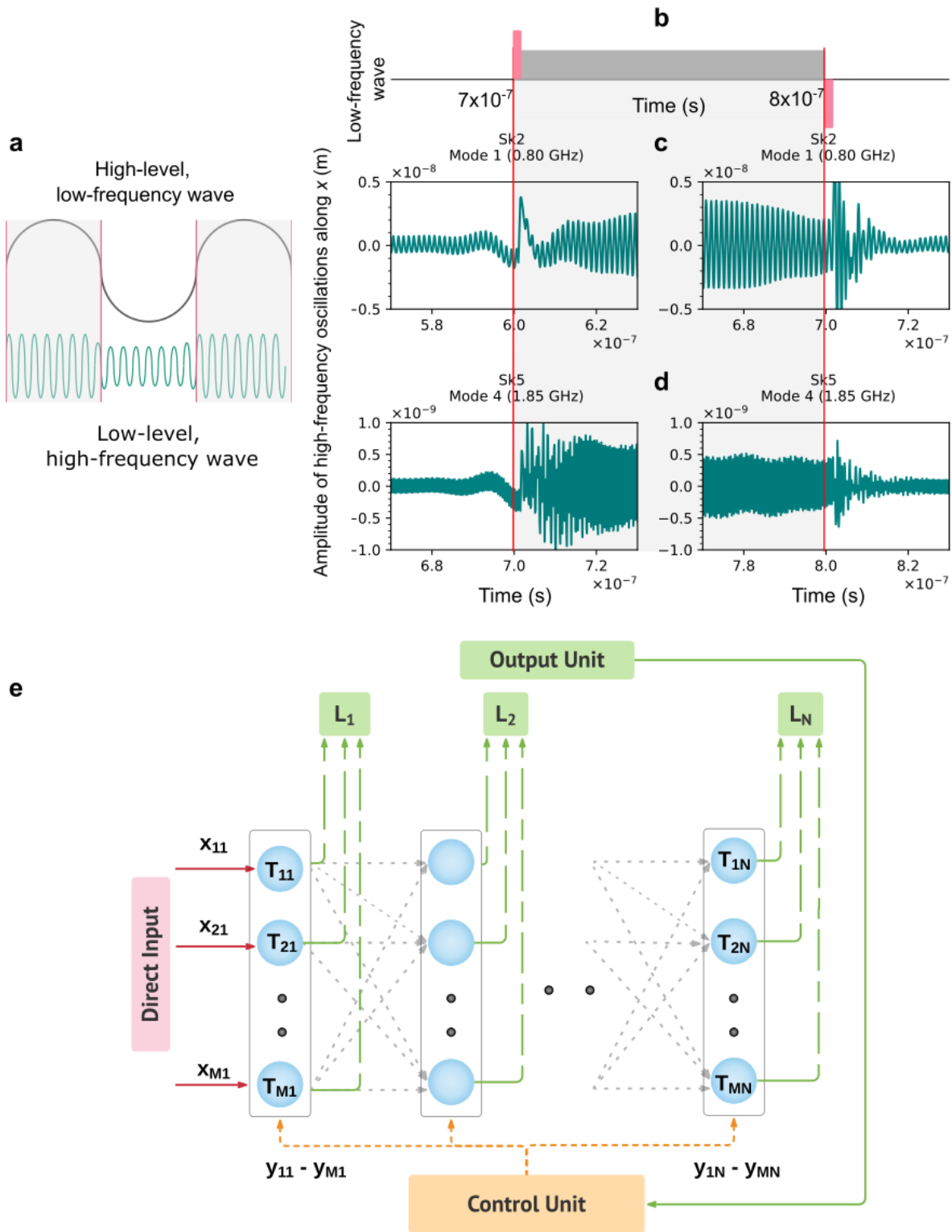

FIG. 6. CFC and flexible computing using T-SKONE. (a) Schematic of CFC in the brain. (b-d) CFC implemented with T-SKONE. (b) Schematic of the low-frequency modulatory wave (dark gray) comprising the reconfiguring current pulses (red) and fed via the control input of T-SKONE. (c & d) The high-frequency oscillatory output of the neuron as the latter is reconfigured from Configuration I to II to I, for a driving frequency of (c) 0.80 GHz and (d) 1.85 GHz. (e) Design of a structurally flexible ANN using T-SKONE with N layers comprising M neurons each. Direct input ($x_{ij}$), control input ($y_{ij}$) and output signals are shown in magenta, orange, and green lines respectively. The control input can selectively switch every neuron ($T_{ij}$) on or off, changing the topology of the network. (a & e) is created in Lucidchart [126].

The proposed ANN is feed-forward, fully connected and consists of *N* layers, with each layer comprising *M* T-SKONEs. A neuron ($T_{ij}$) receives a direct input ($x_{ij}$), shown in Fig. 6 (e) in gray lines, encoded as above in a sinusoidal magnetic field with frequency 0.80 GHz and carrying information from the dataset or the synaptic layer. The neuron also receives a control input ($y_{ij}$), shown in Fig. 6 (e) in orange lines, encoded in modulatory currents that reconfigure T-SKONE and alter its state. The control input $y_{ij} = 0$ (1) corresponds to $T_{ij}$ in Configuration I (II) with its state off (on). Thus, the control signals can be selected to vary the number of hidden layers from 0 to N-2 and vary the number of neurons in every layer from 1 to M. Input to the ANN is fed to the first layer (plotted with magenta lines in Fig. 6 (e)) and outputs from all neurons are sent to the output unit. The effective output layer of the ANN is the right-most layer that gives non-zero outputs.

Network topology can be selected by pre-defined execution-control policies [101] or can be dynamically tuned by a control unit (shown in orange in Fig. 6 (e)) which monitors the output of the network and varies network topology by selecting the control signals ($y_{ij}$). The network can be trained end-to-end by the stochastic gradient descent method or trained layer by layer. Both techniques of growing and pruning [100] can be implemented. To implement growing, the network is initialized to a minimal layer configuration, trained, tested and additional layers are added to enhance the accuracy. To implement pruning, the network is initialized to the largest size and is progressively pruned by the removal of neurons until the desired accuracy is reached with the smallest network size.

**Fundamental benefits of T-SKONE-based reconfigurable ANNs -**

*1. General AI* - Flexible networks provide a general problem-solving tool that can be reconfigured to implement a variety of tasks [66].

*2. Compact architecture and energy-efficiency* – Flexible networks can be pruned to minimize the network size [100], making the ANN compact and energy-efficient.

*3. Fault-tolerant* - Flexible networks can be easily tested, and their faults can be isolated [66].

Therefore, structurally flexible networks built using T-SKONE can mitigate multiple challenges facing conventional ANNs including their lack of general intelligence [24] and their large energy and area footprint [78].

**Demonstration of feature-binding using T-SKONE** – Feature binding enables the brain to combine different features of a perceived object into a coherent entity (see Fig. 7 (a)) and integrate various fragments of sensory information to build a coherent representation of the outside world [15]. Feature binding is known to be important for visual cognition [22] and is a related to consciousness [23]. Taking inspiration from the brain [16, 17, 18, 19, 20, 21], here feature binding is implemented by feeding a low-frequency wave into the control input of T-SKONE such that the phase of this wave modulates the frequency of the high-frequency neuronal output.

In this task, a network of T-SKONEs (see Fig. 7 (b)) receives visual information about two objects, a 'red circle' and a 'blue cross', in sequence and 100 ns apart. The two object features are color and shape. The ANN comprises a single layer of 10 T-SKONEs ($T_i$) that process information in parallel. Through the direct input ($x_i$) encoded in the amplitude of an oscillating magnetic field, one neuron ($T_1$) is fed information about the object color, while nine neurons ($T_2$-$T_{10}$) are fed information about the object shape (see Fig. 7 (b)). The control input ($y_i$) is a low-frequency modulatory wave of 10 MHz. Micromagnetic simulations show that as the phase of the modulatory wave changes from 0 to $\pi$, all neurons are reconfigured from Configuration I to II (see Fig. 7 (c)), which changes the frequencies at which the perceived features are encoded. Details about the simulations are given in Methods.

The 'red circle' which is perceived first, is encoded by the ANN primarily in frequency $f_0$ (1.22 GHz) as the neurons are in Configuration I at the time of perception (see Fig. 7 (d - e)). Alternatively, the 'blue cross' which is perceived 100 ns later, is encoded to a greater extent in frequency $f_1$ (1.85 GHz) than $f_0$ as the neurons are in Configuration II (see Fig. 7 (f - g)). Therefore, the features of the same object are encoded at the same frequency, while features of different objects are encoded at different frequencies by the ANN. The object features can be automatically combined in hardware via synchronization to realize feature binding. Using additional circuitry to filter out the frequency component with a smaller amplitude will encode the first object purely at $f_0$ and the second object purely at $f_1$, making the ANN more robust.

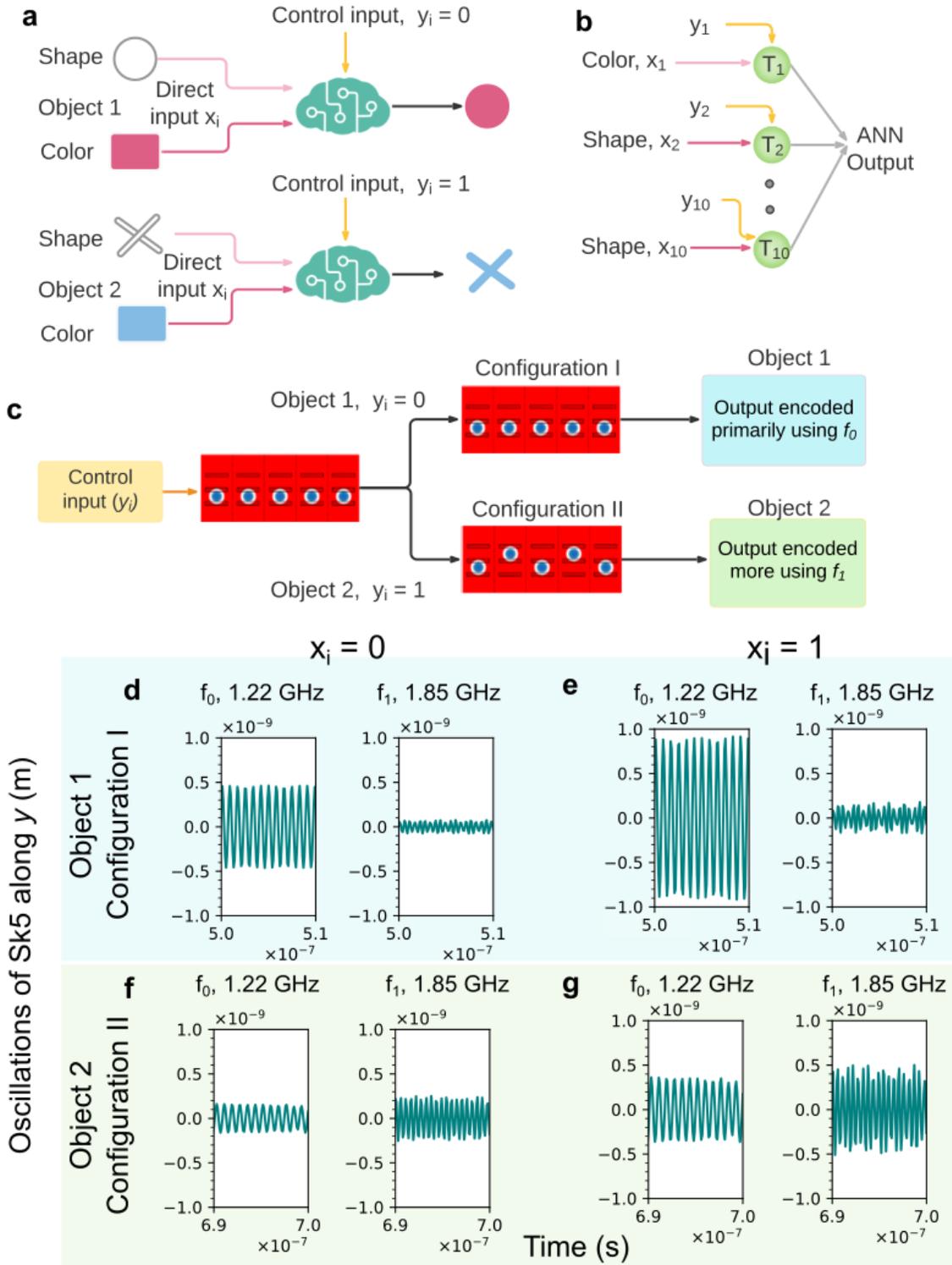

FIG. 8. Neuromorphic implementation of feature binding. (a) Schematic of the feature binding. An intelligent agent combines two features, shape and color, to form a coherent percept. (b) Diagram of the adaptive ANN which consists of 1 layer of 10 neurons ($T_i$). Object features are encoded in direct input ($x_i$), with color encoded in $x_1$ and shape in $x_2$-$x_{10}$. (c) Flowchart of the feature binding process. (d-g) Plot of the oscillations of $Sk5$ along $y$ with the direct input ($x_i$) carried by the sum of two sinusoidal magnetic fields of frequencies $f_0$ (1.22 GHz) and $f_1$ (1.85 GHz). Each subplot shows the neuronal output resolved into frequency components $f_0$ and $f_1$. Input $x_i = 0$ is encoded by a magnetic field amplitude of 2.5 $Oe$ and $x_i = 1$ is encoded by an amplitude of 5.0 $Oe$. (d & e) Neuronal output for object 1 ($y_i = 0$); the neuron is in Configuration I and the output is predominantly encoded in frequency $f_0$. (f & g) Neuronal output for object 2 ($y_i = 1$); the neuron is in Configuration II and the output is encoded to a greater extent in frequency $f_1$.

## Discussion

In conclusion, this work proposes the design of a previously unachieved, hardware-based adaptive neuron. This neuron is subsequently utilized to realize three cognitive abilities, namely, context-awareness, cross-frequency coupling and feature binding. Context-awareness and cross-frequency coupling have not yet been realized in hardware-based neurons. Additionally, the proposed neuron is used to construct an adaptive neural network and perform context-aware diagnosis of breast cancer. Our simulations show that the adaptive ANN achieves diagnosis with higher accuracy while learning faster from a smaller dataset and using a more compact and energy-efficient network than the state-of-the-art ANNs that use non-adaptive neurons. The work further describes how hardware-based adaptive neurons like T-SKONE can mitigate several important challenges facing contemporary ANNs by enabling faster learning from smaller datasets, compact architectures, general AI applications, energy-efficiency and fault-tolerance. However, realizing the potential of T-SKONE requires further development of spin-wave based synapses and methods for wafer-scale integration. Additionally, designing adaptive neurons where contextual inputs can acquire continuous values and the non-linearity of the transfer function can be tuned over a wide range will be significant steps forward.

## Methods

**Material parameters for micromagnetic modeling** - Material parameters used for micromagnetic modeling of the TmIG/Pt system were taken from references [84] [102]: saturation magnetization $M_{sat} = 50 \times 10^3 \, A/m$, exchange stiffness $A_{ex} = 0.8 \times 10^{-12} \, J/m$, interfacial Dzyaloshinskii-Moriya constant $D_{ind} = 50 \times 10^{-6} \, J/m^2$, Gilbert damping constant $\alpha = 0.02$, perpendicular magnetic anisotropy (PMA) $K_u = 1.8 \times 10^3 \, J/m^3$ and $2.0 \times 10^3 \, J/m^3$ for the reduced anisotropy regions (pinning sites) and normal regions, respectively. The neuron consists of five nanotracks of width of 180 nm (along *x*, coordinates defined in Fig. 1). There were two high-anisotropy track regions of length 56 nm (along *y*) alternating between three low-anisotropy pinning wells of length 12 nm (along *y*). The neuron was padded along *x* and *y* with unpatterned magnetic material to enhance skyrmion stability and closed boundary conditions were employed. In all simulations, a DC magnetic field of amplitude $H_{DC} = 240 \, Oe$ was applied to the neuron in order to stabilize the skyrmions. The application of this DC field is required for skyrmion stability specifically in the TmIG/Pt system [84], but is not required for the neuron's design or functionality.

**Simulation of the FFT spectrum** - To calculate the FFT spectrum, the neuron was first excited with an external magnetic field $H_{RF} = H_0 sinc(\omega t)$ along *x* for 1 ns, with the amplitude of the magnetic field $H_0 = 40 \, Oe$ and the cutoff frequency $\omega = 2\pi \times f$ where $f = 60$ GHz. The field was applied to all nanotracks for 1 ns. The neuron was subsequently allowed to relax, and its dynamics during relaxation was examined and the trajectories of the skyrmion cores were calculated using the method described in [52].

**Amplitude Modulation** - To demonstrate Amplitude Modulation, T-SKONE was excited in all nanotracks with a sinusoidal magnetic field $H_{RF} = H_0 sin(\omega t)$ along $x$ with $H_0 = 10\ Oe$ and $\omega = 2\pi \times f$, at a range of driving frequencies $f$, and the oscillatory output was calculated. The oscillations of the skyrmion cores along $x$ and $y$ are taken to be the neuronal response.

**Construction of dataset for context-aware breast cancer diagnosis** - In this task, breast cancer diagnosis is performed by an ANN that classifies biopsy image features of breast mass into benign and malignant categories. In addition to direct information that comprises biopsy image features, the ANN also uses contextual information comprising patient medical data to perform the diagnostic prediction.

Owing to the lack of a well-known, multi-modal cancer dataset containing information about both biopsy images and patient medical history, we constructed such a dataset by appending the Breast Cancer Wisconsin (Diagnostic) dataset (comprising biopsy image features) [91] with information regarding patient medical data (comprising attributes $\alpha_i$). The modified dataset was constructed so as to capture the statistical correlations reported in literature between the attributes or risk factors and the presence of cancer. For instance, the dataset replicated the real-world situation where breast cancer was 100 times more likely for women than for men. This dataset was obtained by using Bayesian inference, the details of which are described in Supplementary Information S4. It is worthwhile to note that in real-world applications, information about attributes for a patient should be easily available from their medical history. This method can be extended to predict malignancy from mammographic elements for early cancer screening [103].

Patient medical data comprises well-known risk factors for breast cancer and was used as contextual input that provides enhanced information about a patient's health to the neural network. Four risk factors for breast cancer were selected. For every factor (labeled attribute $\alpha_i$), we found information from literature about the risk of getting breast cancer given that a patient has attribute $\alpha_i$ vs. the risk of getting breast cancer if that patient does not have attribute $\alpha_i$. This is labeled relative risk ($u_i$). The attributes selected for this study are described below.

i. **Patient is female** - The risk for breast cancer in America is 100 times higher if the patient is female rather than male [65].
ii. **Age of the patient is greater than 50 years** - Age is one of the most important risk factors of breast cancer, as the risk for developing cancer is known to be 1/53 for age < 49 years, 1/43 for 50< age <59 years, 1/23 for 60 < age <69 and 1/15 for age >70 across the world [104].
iii. **Patient has a Body Mass Index (BMI) in the obese range** - Women with a BMI in the obese range (BMI > 30) have a threefold increased risk of breast cancer, as reported by a study conducted in Iran [105].

iv. **Alcohol intake of patient is high** – In data collected worldwide, women with a high alcohol intake (of at least 27 units per week) were 51% more likely to develop breast cancer as compared to non-drinkers [65].

These four binary attributes were treated as independent variables and formed the four contextual inputs, such that every attribute (e.g. patient is female) was associated with 'True' or 'False' values. This data was fed to the ANN as contextual information to perform the cancer diagnosis.

**Micromagnetic simulations to demonstrate feature-binding** - Feature binding is a cognitive ability of the brain by which it combines different features of a perceived object that are processed in different regions of the brain to form a coherent perception of the entire object. Taking inspiration from the brain, we implement feature binding based on synchrony and phase-coding. Therein, the phase of a low-frequency modulatory oscillation helps to bind the various features of an object by ensuring that different features of the same object are encoded at the same frequency, such that they can be correctly combined through synchronization.

In this task, a network of 10 T-SKONEs was designed to receive visual information about two objects, a 'red circle' and a 'blue cross', in sequence and 100 ns apart. The two object features are color and shape. The ANN was designed as a single layer network of 10 T-SKONEs ($T_i$) such that one neuron ($T_1$) processes information about the object color, while nine neurons ($T_2$-$T_{10}$) processes information about the object shape, in parallel.

Information about the object features was fed as direct input ($x_i$) to the neurons and encoded in the amplitude of an oscillating magnetic field $H_{RF,i} = H_{0,i}(sin(\omega_0 t) + sin(\omega_1 t))$, where $\omega_j = 2\pi \times f_j$ and $f_0 = 1.22$ GHz and $f_1 = 1.85$ GHz. The color 'red' ('blue') corresponded to $x_1 = 0$ (1) and was input to neuron $T_1$. The shape of the objects was in the form of a 3×3 pixel array, where 'off' ('on') pixel states corresponded to $x_i = 0$ (1), for i = 2-10. Thus, the 'circle' was represented with an input vector {$x_2, x_3,.., x_{10}$} = {1, 1, 1, 1, 0, 1, 1, 1, 1}, and the 'cross' was represented with a vector {$x_2, x_3,.., x_{10}$} = {1, 0, 1, 0, 1, 0, 1, 0, 1}. The direct input for every neuron, $x_i = 0$ (1) corresponded to magnetic field amplitude $H_{0,i} = 2.5\ Oe$ (5.0 $Oe$). In a hardware implementation of this task, the visual information regarding object features could perceived by a retinomorphic sensor [106, 107] and converted by a waveform generator into an oscillating magnetic field which would be fed as direct input ($x_i$) to the T-SKONE layer.

The control input ($y_i$) was a low-frequency modulatory wave of 10 MHz which constituted modulatory currents $J_2$ and $J_4$, such that all neurons ($T_i$) switched their configurations every 100 ns, thereby switching their output frequencies. Thus, objects perceived in different time windows were encoded at different frequencies. Micromagnetic simulations were conducted for a T-SKONE in the ANN described above, and the effective output of the neuron was taken to be the oscillatory output of *Sk5* along *y*. This oscillatory output was filtered to obtain the two frequency components, i.e., $f_0 = 1.22$ GHz and $f_1 = 1.85$ GHz.

# References

[1] E. Neftci, J. Binas, U. Rutishauser, E. Chicca, G. Indiveri and R. J. Douglas, "Synthesizing cognition in neuromorphic electronic systems," *Proceedings of the National Academy of Sciences, USA,* vol. 110, no. 37, pp. E3468-E3476, 2013.

[2] S. Wen, A. Rios, Y. Ge and L. Itti, "Beneficial Perturbation Network for Designing General Adaptive Artificial Intelligence Systems," 2021.

[3] S. Grossberg, "Toward Autonomous Adaptive Intelligence: Building Upon Neural Models of How Brains Make Minds," *IEEE Transactions on Systems, Man, and Cybernetics: Systems,* vol. 51, no. 1, pp. 51-75, 2021.

[4] T.-J. Huang, "Imitating the Brain with Neurocomputer A "New" Way Towards Artificial General Intelligence," *International Journal of Automation and Computing,* vol. 14, no. 5, pp. 520-531, 2017.

[5] P. Dayan, "Simple substrates for complex cognition," *Frontiers of Neuroscience,* vol. 2, no. 2, p. 255–263, 2008.

[6] "Computational intelligence framework for context-aware decision making," *International Journal of System Assurance Engineering and Management volume,* vol. 8, p. 2146–2157, 2017.

[7] A. K. dey, "Understanding and Using Context," *Personal and Ubiquitous Computing,* vol. 5, pp. 4-7, 2001.

[8] B. Schilit and M. Theimer, "Disseminating active map information to mobile hosts," *IEEE Network,* vol. 8, pp. 22-32, 1994.

[9] M. A. Whittington, R. D. Traub and N. E. Adams, "A future for neuronal oscillation research," *Brain and Neuroscience Advances,* vol. 2, pp. 1-6, 2018.

[10] O. Jensen and L. Colgin, "Cross-frequency coupling between neuronal oscillations," *Trends in Cognitive Sciences,* vol. 11, no. 7, pp. 267-269, 2007.

[11] J. E. Lisman and O. Jensen, "The theta–gamma neural code," *Neuron,* vol. 77, p. 1002–1016, 2013.

[12] A.-L. Giraud and D. Poeppel, "Cortical oscillations and speech processing: emerging computational principles and operations," *Nature Neuroscience,* vol. 15, p. 511–517, 2012.

[13] O. Jensen, M. Bonnefond and R. VanRullen, "An oscillatory mechanism for prioritizing salient unattended stimuli," *Trends in Cognitive Sciences,* vol. 16, no. 4, p. 200–206, 2012.

[14] N. Axmacher, M. M. Henseler, O. Jensen, I. Weinreich, C. E. Elger and J. Fell, "Cross-frequency coupling supports multi-itemworking memory in the human hippocampus," *Proceedings of the National Academy of Sciences of the USA,* vol. 107, no. 7, p. 3228–3233, 2010.

[15] C. Wan, P. Cai, X. Guo, M. Wang, N. Matsuhisa, L. Yang, Z. Lv, Y. Luo, X. J. Loh and X. Chen, "An artificial sensory neuron with visual-haptic fusion," *Nature Communications,* vol. 11, no. 4602, 2020.

[16] N. Hakim and E. K. Vogel, "Phase-coding memories in mind," *PLOS Biology,* vol. 16, no. 8, p. 1, 2018.

[17] A. J. Watrous, J. Miller, S. E. Qasim, I. Fried and J. Jacobs, "Phase-tuned neuronal firing encodes human contextual representations for navigational goals," *eLife,* vol. 7, p. e32554, 2018.

[18] M. Volgushev, M. Chistiakova and W. Singer, "Modification of discharge patterns of neocortical neurons by induced oscillations of the membrane potential," *Neuroscience,* vol. 83, no. 1, pp. 15-25, 1998.

[19] P. Fries, D. Nikolić and W. Singer, "The gamma cycle," *Trends in Neuroscience,* vol. 30, no. 7, pp. 309-316, 2007.

[20] M. X. Cohen, "Fluctuations in oscillation frequency control spike timing and coordinate neural networks," *Journal of Neuroscience,* vol. 34, no. 27, p. 8988–8998, 2014.

[21] C. Kayser, R. A. Ince and S. Panzeri, "Analysis of slow (theta) oscillations as a potential temporal reference frame for information coding in sensory cortices," *PLoS Computational Biology,* vol. 8, no. 10, p. e1002717, 2012.

[22] J. Saiki, "Feature binding in object-file representations of multiple moving items," *Journal of Vision,* vol. 3, no. 2, pp. 6-21, 2003.

[23] S. Zmigrod and B. Hommel, "The relationship between feature binding and consciousness: Evidence from asynchronous multi-modal stimuli," *Consciousness and Cognition,* vol. 20, p. 586, 2011.

[24] DARPA, "DARPA Perspectives on AI," [Online]. Available: https://www.darpa.mil/about-us/darpa-perspective-on-ai.

[25] B. Chatterjee, N. Cao, A. Raychowdhury and S. Sen, "Context-Aware Intelligence in Resource-Constrained IoT Nodes: Opportunities and Challenges," *IEEE Design & Test,* vol. 36, no. 2, pp. 7-40, 2019.

[26] E. R. Kandel, S. Mack, T. M. Jessell, J. H. Schwartz, S. A. Siegelbaum and A. J. Hudspeth, Principles of Neural Science, McGraw Hill Professional, 2013.

[27] T. L. Fonseca and L. Goliatt, "Hybrid Extreme Learning Machine and Backpropagation with Adaptive Activation Functions for Classification Problems," in *International Conference on Intelligent Systems Design and Applications*, 2020.

[28] M. M. Lau and K. Hann Lim, "Review of Adaptive Activation Function in Deep Neural Network," 2018.

[29] F. Nadim and D. Bucher, "Neuromodulation of Neurons and Synapses," *Curr Opin Neurobiol.,* vol. 0, p. 48–56, 2014.

[30] E. Marder, "Dynamic Modulation of Neurons and Networks," in *Advances in Neural Information Processing Systems 6 (NIPS)*, 1993.

[31] M. Siegel, A. K. Engel and T. H. Donner, "Cortical network dynamics of perceptual decision-making in the human brain," *Frontiers in Human Neuroscience,* vol. 5, p. 21, 2011.

[32] R. F. Helfrich and R. T. Knight, "Oscillatory Dynamics of Prefrontal Cognitive Control," *Trends in Cognitive Sciences,* vol. 20, no. 12, p. 916, 2016.

[33] C. Tallon-Baudry and O. Bertrand, "Oscillatory gamma activity in humans and its role in object representation," *Trends in Cognitive Sciences,* vol. 3, no. 4, p. 151, 1999.

[34] W. Singer and C. M. Gray, "Visual feature integration and the temporal correlation hypothesis," *Annu. Rev. Neurosci.,* vol. 18, p. 555, 1995.

[35] S. Karakas, C. Basar-Eroglu, C. Ozesmi, H. Kafadar and O. U. Erzengin, "Gamma response of the brain: a multifunctional oscillation that represents bottom-up with top-down processing," *International Journal of Psychophysiology,* vol. 39, pp. 137-150, 2001.

[36] A. Wutz, D. Melcher and J. Samaha, "Frequency modulation of neural oscillations according to visual task demands," *PNAS,* vol. 115, no. 6, p. 1346–1351, 2018.

[37] B. Voytek, R. T. Canolty, A. Shestyuk, N. E. Crone, J. Parvizi and R. T. Knight, "Shifts in gamma phase–amplitude coupling frequency from theta to alpha over posterior cortex during visual tasks," *Frontiers in Human Neuroscience,* vol. 4, p. 191, 2010.

[38] G. Indiveri, "Modeling Selective Attention Using a Neuromorphic Analog VLSI Device," *Neural Computation,* vol. 12, p. 2857–2880, 2000.

[39] K. Hornik, M. Stinchcombe and H. White, "Multilayer Feedforward Networks are Universal Approximators," *Neural Networks,* vol. 2, pp. 359-366, 1989.

[40] A. N. Kolmogorov, "On the representation of continuous functions of many variables by superposition of continuous functions of one variable and addition," *Doklady Akademii Nauk,* vol. 114, pp. 953-956, 1957.

[41] F. Piazza, A. Uncini and M. Zenobi, "Artificial neural networks with adaptive poly- nomial activation function," in *International Joint Conference on Neural Networks*, Beijing, 1992.

[42] G. Tezel and Y. Özbay, "A New Neural Network with Adaptive Activation Function for Classification of ECG Arrhythmias," in *International Conference on Knowledge-Based and Intelligent Information and Engineering Systems*, 2007.

[43] V. Kunc and J. Kléma, "On transformative adaptive activation functions in neural networks for gene expression inference," *PLoS ONE,* vol. 16, no. 1, p. e0243915, 2021.

[44] M. Davies et al., "Loihi: A Neuromorphic Manycore Processor with On-Chip Learning," in *IEEE Micro*, 2018.

[45] I. S. Jones and K. Kording, "Can Single Neurons Solve MNIST? The Computational Power of Biological Dendritic Trees," *arXiv:2009.01269.*

[46] C. Koch and I. Segev, "The role of single neurons ininformation processing," *Nature Neuroscience,* vol. 3, p. 1171, 2000.

[47] A. Shaban, S. S. Bezugam and M. Suri, "An adaptive threshold neuron for recurrent spiking neural networks with nanodevice hardware implementation," *Nature Communications,* vol. 12, p. 4234, 2021.

[48] L. Gao, P.-Y. Chen and S. Yu, "NbOx based oscillation neuron for neuromorphic computing," *Appl. Phys. Lett.,* vol. 111, p. 103503, 2017.

[49] T. Jackson, S. Pagliarini and L. Pileggi, "An Oscillatory Neural Network with Programmable Resistive Synapses," in *28 nm CMOS, IEEE International Conference on Rebooting Computing*, 2018.

[50] M. Ignatov, M. Ziegler, M. Hansen and H. Kohlstedt, "Memristive stochastic plasticity enables mimicking of neural synchrony: Memristive circuit emulates an optical illusion," *Science Advances,* vol. 3, p. e1700849, 2017.

[51] A. Vansteenkiste, J. Leliaert, M. Dvornik, M. Helsen, F. Garcia-Sanchez and B. Van Waeyenberge, "The design and verification of MuMax3," *AIP advances,* vol. 4, no. 10, p. 107133, 2014.

[52] J. Kim, J. Yang, Y.-J. Cho, B. Kim and S.-K. Kim, "Coupled gyration modes in onedimensional skyrmion arrays in thin-film nanostrips as new type of information carrier," *Scientific Reports,* vol. 7, p. 45185, 2017.

[53] M. I. Trukhanova and P. Andreev, "A quantum hydrodynamical model of skyrmions with electrical dipole moments and novel magneto-electric skyrmion Hall effect," *Progress of Theoretical and Experimental Physics,* vol. 2020, no. 4, p. 043I01, 2020.

[54] J. Grollier, D. Querlioz, K. Y. Camsari, K. Everschor-Sitte, S. Fukami and M. D. Stiles, "Neuromorphic spintronics," *Nature Electronics,* vol. 3, p. 360–370, 2020.

[55] S. Singh, A. Sarma, N. Jao, A. Pattnaik, S. Lu, K. Yang, A. Sengupta, V. Narayanan and C. R. Das, "NEBULA: A Neuromorphic Spin-Based Ultra-Low Power Architecture for SNNs and ANNs," in *ACM/IEEE 47th Annual International Symposium on Computer Architecture (ISCA)*, Valencia, Spain, 2020.

[56] W. Kang, Y. Huang, X. Zhang, Y. Zhou and W. Zhao, "Skyrmion-Electronics: An Overview and Outlook," *Proceedings of the IEEE,* vol. 104, no. 10, pp. 2040 - 2061, 2016.

[57] D. Kuzum, R. G. D. Jeyasingh, B. Lee and H.-S. P. Wong, "Nanoelectronic Programmable Synapses Based on Phase Change Materials for Brain-Inspired Computing," *Nano Letters,* vol. 12, no. 5, p. 2179–2186, 2012.

[58] Z. Wang, S. Khandelwal and A. I. Khan, "Ferroelectric Oscillators and Their Coupled Networks," *IEEE Electron Device Letters,* vol. 38, no. 11, pp. 1614 - 1617, 2017.

[59] S. Liu, C. H. Bennett, J. S. Friedman, M. J. Marinella, D. Paydarfar and J. A. C. Incorvia, "Controllable reset behavior in domain wall-magnetic tunnel junction artificial neurons for task-adaptable computation," *IEEE Magnetics Letters,* vol. 12, p. 4500805, 2021.

[60] S. Liu, T. P. Xiao, C. Cui, J. A. C. Incorvia, C. H. Bennett and M. J. Marinella, "A domain wall-magnetic tunnel junction artificial synapse with notched geometry for accurate and efficient training of deep neural networks," *Applied Physics Letters,* vol. 118, p. 202405, 2021.

[61] C. Cui, O. G. Akinola, N. Hassan, C. H. Bennett, M. J. Marinella, J. S. Friedman and J. A. C. Incorvia, "Maximized lateral inhibition in paired magnetic domain wall racetracks for neuromorphic computing," *IOP Nanotechnology,* vol. 31, p. 29, 2020.

[62] M. Alamdar et al., "Domain wall-magnetic tunnel junction spin orbit torque device and circuit prototypes for in-memory computing," *Applied Physics Letters,* vol. 118, p. 112401, 2021.

[63] R. Yan, et al., "Richer fusion network for breast cancer classification based on multimodal data," *BMC Medical Informatics and Decision Making,* vol. 21, no. 1, p. 134, 2021.

[64] D. Sun, M. Wang and A. Li, "A Multimodal Deep Neural Network for Human Breast Cancer Prognosis Prediction by Integrating Multi-Dimensional Data," *IEEE/ACM Transactions on Computational Biology and Bioinformatics,* vol. 16, no. 3, p. 841, 2019.

[65] Y.-S. Sun, et al., "Risk Factors and Preventions of Breast Cancer," *International Journal of Biological Sciences,* vol. 13, no. 11, pp. 1387-1397, 2017.

[66] S. Satyanarayana, Y. P. Tsividis and H. P. Graf, "A Reconfigurable VLSI Neural Network," *IEEE Journal of Solid-State Circuits,* vol. 27, no. 1, pp. 67-81, 1992.

[67] A. A. Abdellatif, A. Mohamed, C. F. Chiasserini, M. Tlili and A. Erbad, "Edge Computing for Smart Health: Context-Aware Approaches, Opportunities, and Challenges," *IEEE Network,* vol. 33, no. 3, pp. 196-203, 2019.

[68] R. Fernandez-Rojas, A. Perry, H. Singh, B. Campbell, S. Elsayed, R. Hunjet and H. A. Abbass, "Contextual Awareness in Human-Advanced-Vehicle Systems: A Survey," *IEEEAccess,* vol. 7, pp. 33304-33328, 2019.

[69] F. Belkadi, M. A. Dhuieb, J. V. Aguado, F. Laroche, A. Bernard and F. Chinesta, "Intelligent assistant system as a context-aware decision-making support for the workers of the future," *Computers & Industrial Engineering,* vol. 139, p. 105732, 2020.

[70] O. B. Sezer, E. Dogdu and A. M. Ozbayoglu, "Context-Aware Computing, Learning, and Big Data in Internet of Things: A Survey," *IEEE Internet of Things Journal,* vol. 5, no. 1, 2018.

[71] T. Rausch and S. Dustdar, "Edge Intelligence: The Convergence of Humans, Things, and AI," in *2019 IEEE International Conference on Cloud Engineering (IC2E)*, 2019.

[72] E. Ceolini, C. Frenkel, S. B. Shrestha, G. Taverni, L. Khacef, M. Payvand and E. Donati, "Hand-Gesture Recognition Based on EMG and Event-Based Camera Sensor Fusion: A Benchmark in Neuromorphic Computing," *Frontiers in Neuroscience,* vol. 14, p. 637, 2020.

[73] J. Fayyad, M. A. Jaradat, D. Gruyer and H. Najjaran, "Deep Learning Sensor Fusion for Autonomous Vehicle Perception and Localization: A Review," *Sensors,* vol. 20, no. 15, p. 4220, 2020.

[74] M. Muzammal, R. Talat, A. H. Sodhro and S. Pirbhulal, "A multi-sensor data fusion enabled ensemble approach for medical data from body sensor networks," *Information Fusion,* vol. 53, pp. 155-164, 2020.

[75] R. R. Shamshiri, I. Bojic, E. van Henten, S. K. Balasundram, V. Dworak, M. Sultan and C. Weltzien, "Model-based evaluation of greenhouse microclimate using IoT-Sensor data fusion for energy efficient crop production," *Journal of Cleaner Production,* vol. 263, p. 121303, 2020.

[76] N. C. Thompson, K. Greenewald, K. Lee and G. Manso, "The Computational Limits of Deep Learning," *https://arxiv.org/abs/2007.05558*.

[77] D. Kahneman, Thinking Fast and Slow, Penguin Books, 2011.

[78] Cisco, "Cisco Global Cloud Index: Forecast and methodology 2016–2021 (Cisco, document 1513879861264127)," 2018.

[79] L. N. Long and C. F. Cotner, "A Review and Proposed Framework for Artificial General Intelligence," in *IEEE Aerospace Conference*, 2019.

[80] J. Wells, J. H. Lee, R. Mansell, R. P. Cowburn and O. Kazakova, "Controlled manipulation of domain walls in ultra-thin CoFeB nanodevices," *Journal of Magnetism and Magnetic Materials,* vol. 400, pp. 219-224, 2016.

[81] K. Moon, B. Chun, W. Kim and C. Hwang, "Control of spin-wave refraction using arrays of skyrmions," *Physical Review Applied,* vol. 6, no. 6, p. 064027, 2016.

[82] M. Mochizuki, "Spin-Wave Modes and Their Intense Excitation Effects in Skyrmion Crystals," *Phys. Rev. Lett. ,* vol. 108, p. 017601, 2012.

[83] Y. Onose, Y. Okamura, S. Seki, S. Ishiwata and Y. Tokura, "Observation of Magnetic Excitations of Skyrmion Crystal in a Helimagnetic Insulator $Cu_2OSeO_3$," *Phys. Rev. Lett. ,* vol. 109, p. 037603, 2012.

[84] Q. Shao, Y. Liu, G. Yu, S. Kim, X. Che, C. Tang, Q. He, Y. Tserkovnyak, J. Shi and K. Wang, "Topological Hall effect at above room temperature in heterostructures composed of a magnetic insulator and a heavy metal," *Nature Electronics,* vol. 2, no. 5, pp. 182-186., 2019.

[85] S. Vélez, J. Schaab, M. Wörnle, M. Müller, E. Gradauskaite, P. Welter, C. Gutgsell, C. Nistor, C. Degen, M. Trassin and M. Fiebig, "High-speed domain wall racetracks in a magnetic insulator," *Nature communications,* vol. 10, no. 1, pp. 1-8, 2019.

[86] C. Avci, E. Rosenberg, L. Caretta, F. Büttner, M. Mann, C. Marcus, D. Bono, C. Ross and G. Beach, "Interface-driven chiral magnetism and current-driven domain walls in insulating magnetic garnets," *Nature nanotechnology,,* vol. 14, no. 6, pp. 561-5, 2019.

[87] M. Romera, et al., "Vowel recognition with four coupled spin-torque nano-oscillators," *Nature,* vol. 563, p. 230, 2018.

[88] A. Khitun, D. E. Nikonov, M. Bao, K. Galatsis and K. L. Wang, "Feasibility study of logic circuits with a spin wave bus," *Nanotechnology,* vol. 18, no. 46, p. 465202, 2007.

[89] D. Bhowmin et al., "On-chip learning for domain wall synapse based Fully Connected Neural Network," *Journal of Magnetism and Magnetic Materials,* vol. 489, p. 165434, 2019.

[90] A. K. Dey, D. Salber, G. D. Abowd and M. Futakawa, "Providing Architectural Support for Building Context-Aware Application," College of Computing, Georgia Institute of Technology, 2000.

[91] O. L. Mangasarian and W. H. Wolberg, "Cancer diagnosis via linear programming," *SIAM News,* vol. 23, no. 5, pp. 1, 18, 1990.

[92] N. Ali et al., "Automatic label-free detection of breast cancer using nonlinear multimodal imaging and the convolutional neural network ResNet50," *Translational Biophotonics,* vol. 1, no. 1-2, p. e201900003, 2019.

[93] D. P. Kingma and J. Ba, "A method for stochastic optimization," *arXiv:1412.6980* .

[94] A. Paszke, et al., "Pytorch: An imperative style, high-performance deep learning library," *Advances in Neural Information Processing Systems ,* vol. 32, pp. 8024 - 8035, 2019.

[95] R. Yan et al., "Integration of Multimodal Data for Breast Cancer Classification Using a Hybrid Deep Learning Method," in *Intelligent Computing Theories and Application. ICIC*, 2019.

[96] P. Smolensky, "Next-generation architectures bridge gap between neural and symbolic representations with neural symbols," Microsoft Research, 12 December 2019. [Online]. Available: https://www.microsoft.com/en-us/research/blog/next-generation-architectures-bridge-gap-between-neural-and-symbolic-representations-with-neural-symbols/.

[97] K. Yi, et al., "CLEVRER: Collision events for video representation and reasoning," in *ICLR*, 2020.

[98] L. Zhaoping, "A new framework for understanding vision from the perspective of the primary visual cortex," *Current Opinion in Neurobiology,* vol. 58, pp. 1-10, 2019.

[99] A. E. Sulavko, D. A. Volkov, S. Zhumazhanova and R. V. Borisov, "Subjects Authentication Based on Secret Biometric Patterns Using Wavelet Analysis and Flexible Neural Networks," in *2018 XIV International Scientific-Technical Conference on Actual Problems of Electronics Instrument Engineering (APEIE)*, October 2018.

[100] D. Blalock, J. J. G. Ortiz, J. Frankle and J. Guttag, "What is the state of neural network pruning?," *arXiv:2003.03033,* 2020.

[101] L. Liu and J. Deng, "Dynamic Deep Neural Networks: Optimizing Accuracy-Efficiency Trade-Offs by Selective Execution," *Proceedings of the AAAI Conference on Artificial Intelligence,* vol. 32, no. 1, p. 3675, 2018.

[102] O. Ciubotariu, A. Semisalova, K. Lenz and M. Albrecht, "Strain-induced perpendicular magnetic anisotropy and Gilbert damping of Tm 3 Fe 5 O 12 thin films," *Scientific reports,* vol. 9, no. 1, pp. 1-8, 2019.

[103] C. E. Floyd, et al., "Prediction of Breast Cancer Malignancy Using an Artificial Neural Network," *Cancer,* vol. 74, no. 11, p. 2944, 1994.

[104] A. McGuire, J. A. L. Brown, C. Malone, R. McLaughlin and M. J. Kerin, "Effects of Age on the Detection and Management of Breast Cancer," *Cancers,* vol. 7, no. 2, p. 908, 2015.

[105] N. Khan, F. Afaq and H. Mukhtar, "Lifestyle as risk factor for cancer: Evidence from human studies," *Cancer Letters,* vol. 293, no. 2, pp. 133 - 143, 2010.

[106] P. Lichtsteiner, T. Delbruck and J. Kramer, "Improved ON/OFF temporally differentiating address-event imager," in *Proceedings of the 11th IEEE International Conference on Electronics, Circuits and Systems (ICECS)*, Tel-Aviv, 2004.

[107] A. Vanarse, A. Osseiran and A. Rassau, "A Review of Current Neuromorphic Approaches for Vision, Auditory, and Olfactory Sensors," *Frontiers in Neuroscience,* vol. 10, p. 115, 2016.

[108] I. Goodfellow, Y. Bengio and A. Courville, Deep Learning, The MIT Press, 2016.

[109] Y. LeCun, Y. Bengio and G. Hinton, "Deep learning," *Nature,* vol. 521, p. 436–444, 2015.

[110] S. Pouyanfar, S. Sadiq, Y. Yan, H. Tian, Y. Tao, M. P. Reyes, M.-L. Shyu, S.-C. Chen and S. S. Iyengar, "A Survey on Deep Learning: Algorithms, Techniques, and Applications," *ACM Computing SurveysSeptember,* vol. 51, no. 5, p. 92, 2019.

[111] B. Yoshua, "From System 1 Deep Learning to System 2 Deep Learning," in *NeurIPS*, 11 December 2019.

[112] R. A. Silver, "Neuronal arithmetic," *Nature Reviews,* vol. 11, p. 474, 2010.

[113] S. Li, et al., "Magnetic skyrmion-based artificial neuron device," *Nanotechnology,* vol. 28, no. 31LT01, 2017.

[114] X. Chen, et al., "A compact skyrmionic leaky–integrate–fire spiking neuron device," *Nanoscale,* vol. 10, p. 6139, 2018.

[115] M.-C. Chen, A. Sengupta and K. Roy, "Magnetic Skyrmion as a Spintronic Deep Learning Spiking Neuron Processor," *IEEE Transactions on Magnetics,* vol. 54, p. 1500207, 2018.

[116] D. Sun, M. Wang and A. Li, "A Multimodal Deep Neural Network for Human Breast Cancer Prognosis Prediction by Integrating Multi-Dimensional Data," *IEEE/ACM Transactions on Computational Biology and Bioinformatics,* vol. 16, no. 3, pp. 841-850, 2019.

[117] P. M. A. Milner, "A model for visual shape recognition," *Psychological Review,* vol. 81, no. 6, pp. 521-535, 1974.

[118] S. Grossberg , "Adaptive pattern classification and universal recoding: I. Parallel development and coding of neural feature detectors," *Biological Cybernetics,* vol. 23, p. 121–134, 1976.

[119] C. von der Malsburg, "The correlation theory of brain function," Internal Report. Dept. of Neurobiology, Max-Planck-Institute for Biophysical Chemistry, Göttingen, Germany., 1981.

[120] C. M. Gray and W. Singer, "Stimulus-specific neuronal oscillations in orientation columns of cat visual cortex," *Proc. Natl. Acad. Sci. USA,* vol. 86, pp. 1698-1702, 1989.

[121] C. M. Gray, P. König, A. K. Engel and W. Singer , "Oscillatory responses in cat visual cortex exhibit inter-columnar synchronization which reflects global stimulus properties," *Nature,* vol. 338, pp. 334-37, 1989.

[122] A. Cortese, B. De Martino and M. Kawato, "The neural and cognitive architecture for learning from a small sample," *Current Opinion in Neurobiology,* vol. 55, pp. 33-141, 2019.

[123] J. Gao ,et al., "MGNN: A Multimodal Graph Neural Network for Predicting the Survival of Cancer Patients," in *SIGIR'20*, China, July 25-30, 2020.

[124] A. Tiwari, "A dynamic goal adapted task oriented dialogue agent," *PLOS One,* vol. 16, no. 4, p. e0249030, 2020.

[125] J. Jang, J. Lee, S. Woo, D. J. Sly, L. J. Campbell, J.-H. Cho, S. J. O'Leary, M.-H. Park, S. Han, J.-W. Choi, J. H. Jang and H. Choi, "A microelectromechanical system artificial basilar membrane based on a piezoelectric cantilever array and its characterization using an animal model," *Scientific Reports,* vol. 5, p. 12447, 2015.

[126] Lucidchart. [Online]. Available: https://www.lucidchart.com.

[127] J. Riddle, A. McFerren and F. Frohlich, "Causal role of cross-frequency coupling in distinct components of cognitive control," *Progress in Neurobiology,* vol. 202, p. 102033, 2021.

[128] A. Hyafil, A.-L. Giraud, L. Fontolan and B. Gutkin, "Neural Cross-Frequency Coupling: Connecting Architectures, Mechanisms, and Functions," *Trends in Neuroscience,* vol. 38, no. 11, p. 725, 2015.

[129] O. Jensen and L. L. Colgin, "Cross-frequency coupling between neuronal oscillations Author links open overlay panel," *Trends in Cognitive Sciences,* vol. 11, no. 7, pp. 267-269, 2007.

[130] F.-Y. Tzeng and K.-L. Ma, "Intelligent Feature Extraction and Tracking for Visualizing Large-Scale 4D Flow Simulations," in *Proceedings of the 2005 ACM/IEEE SC|05 Conference*, 2005.

[131] N. M. Mayer, "Critical Echo State Networks that Anticipate Input Using Morphable Transfer Functions," in *Advances in Neural Networks*, 2017.

**Acknowledgments**

**Funding:** The authors would like to thank Prof. Ioannis Mitliagkas for important discussions regarding adaptive neurons in machine learning.

The authors acknowledge funding from the National Science Foundation CAREER under award number 1940788.

The authors also acknowledge computational resources from the Texas Advanced Computing Center (TACC) at the University of Texas at Austin (http://www.tacc.utexas.edu).

**Author contributions:** PJ conceived the idea and designed the project. PJ, CC and SL carried out the calculations. JACI provided technical guidance. All authors discussed the results and wrote the manuscript.